\documentclass[letterpaper]{article}
\usepackage{aaai,times,helvet,courier}
\usepackage{graphicx}
\usepackage{url}
\usepackage[caption=false,font=footnotesize, hangindent=10pt]{subfig}
\usepackage{amsmath,amsfonts,bm,amssymb,amsthm}
\usepackage[linesnumbered,lined,boxed,ruled]{algorithm2e}
\usepackage{comment}
\usepackage{color}

\newcommand{\argmax}{\mathop{\arg\max}}
\newcommand{\SC}{{\cal C}}
\newcommand{\E}[1]{\mathbb{E}[#1]}
\newtheorem{theorem}{{\bf Theorem}}
\def\done{\hspace*{\fill} \rule{1.8mm}{2.5mm} \\}

\title{Social Sensor Placement in Large Scale Networks: A Graph Sampling 
Perspective}

\author{
Junzhou Zhao$^1$,
John C. S. Lui$^2$,
Don Towsley$^3$,
Xiaohong Guan$^1$ and
Pinghui Wang$^2$\\
$^1$MOE KLINNS Lab, Xi'an Jiaotong University\\
$^2$Department of Computer Science \& Engineering, The Chinese University of
Hong Kong\\
$^3$Computer Science Department, University of Massachusetts at Amherst\\
\{jzzhao,phwang,xhguan\}@sei.xjtu.edu.cn, cslui@cse.cuhk.edu.hk,
towsley@cs.umass.edu
}

\begin{document}

\maketitle

\begin{abstract}
Sensor placement for the purpose of detecting/tracking 
news outbreak and preventing 
rumor spreading is a challenging problem in a large scale online social network
(OSN). This problem is a kind of subset selection problem: choosing a small set
of items from a large population so to maximize some prespecified set function.
However, it is known to be NP-complete. Existing heuristics are very costly
especially for modern OSNs which usually contain hundreds of millions of users.
This paper aims to design methods to find \emph{good solutions} that can well
trade off efficiency and accuracy. We first show that it is possible to obtain
a high quality solution with a probabilistic guarantee from a ``{\em candidate 
set}'' of the underlying social network. By exploring this candidate set, one can
increase the efficiency of placing social sensors. We also present how this 
candidate set can be obtained using ``{\em graph sampling}'', 
which has an advantage over previous methods of not requiring 
%which does not require 
the prior knowledge of the complete network topology. Experiments carried out on two 
real datasets demonstrate 
%the effectiveness of this approach.
not only the accuracy and efficiency of our approach, but aslo
effectiveness in detecting and predicting news outbreak.
\end{abstract}

\section{I. Introduction}\label{sec:intro} 

The rising popularity of online social networks (OSNs) has made information
sharing and discovery much easier than ever before. Such social networks shift
the role of participants from few content producers with many consumers to both
producers and consumers of content. While this fundamental change enables
information diversity in the Internet, it introduces the problem of what 
information sources to subscribe to or follow due to users' limited 
attention capacities. For example, journalists need to discover breaking news 
from OSNs in a timely manner, while government officers may want to prevent 
damages caused by destructive riots arose from OSNs (e.g., the England Riots in 
2011), or twitter users may want to track valuable information by following 
limited twitter accounts. Hence, it is important to decide which subset of 
social network accounts to choose as information sources such that the total 
information obtained is maximized.

The task of selecting a small number of accounts (or nodes) to \emph{cover} as 
much valuable information as possible is feasible in modern OSNs because 
messages are shared by users, e.g., a tweet related to breaking news can be 
retweeted, reposted or shared by the followers or friends in social networks. 
Therefore, one only needs to read one participant's tweets and obtain the 
information about the event. Here, we say that diffusible tweets triggered by 
some users and the associated participants form many \emph{information 
cascades}, and the social network accounts which we prefer to select to 
\emph{monitor} are called \emph{social sensors}. Our problem is {\em how to 
select a finite number of social sensors which can discover as many important 
cascades as possible}. 

{\bf Challenges:} Selecting a small set of items from a large population to 
maximize some prespecified set function is a classical combinatorial
optimization problem and its solution has widespread application, e.g, set
cover problem\cite{Khuller1999,Fujito2000}, influence 
maximization\cite{Kempe2003,Li2012} and optimal 
sensing\cite{Leskovec2007,Krause2008}. However, all these works assume that the
complete data (i.e., network topology)
is available in advance. For modern OSNs, the network topology is
usually not available. This is because many of these OSNs have hundreds of
millions of accounts, and OSN service providers usually limit the request
rates. This makes the task of discovering the entire network topology very
difficult, if not impossible. Secondly, even if we know the network topology,
the underlying subset selection problem is NP-complete. When the objective 
function is \emph{submodular}, a greedy algorithm (GA) can find a solution with the
lower bound of $1-1/e\approx 63\%$ to the optimal solution\cite{Nemhauser1978},
and its execution time is polynomial to the size of network. However, the
algorithm does not scale to handle large graphs like modern OSNs
since they usually have hundreds of millions of nodes. The current
state-of-the-art approach is the Accelerated Greedy (AG)\cite{Minoux1978}.
However, it is not guaranteed to be efficient all the time, i.e., in the worst
case, it is as inefficient as the greedy algorithm. Thirdly and most 
importantly, the purpose of placing social sensors is to
capture ``{\em future}'' 
important events, i.e., sensors selected based on historical
data should have good predictive capability of future information
cascades. However, users in
OSNs are highly dynamic, e.g., everyday many new users join in and many 
existing users drop out. This will lead to a poor predictive capability of 
sensors selected based solely on old historical data. To improve sensors' predictive
capability, one has to periodically reselect sensors. 
Therefore, we need a 
{\em computationally efficient} social sensor placement algorithm that
can accurately capture important future events. 

{\bf Proposed Approach:} The above challenges inspire us to develop 
computationally efficient, cheaper (no need to have the topology of an OSN 
beforehand) methods that provide quality guarantees. In this paper, we 
introduce an approach based on 
{\em graph sampling}\cite{LOVASZ1993,Ribeiro2010}. 
The basic idea is that by carefully choosing a set of candidate sensors, we 
can select the final sensors from this candidate set. 
By sampling, the search space 
can be reduced dramatically and efficiency is increased. In this study, we 
show that graph sampling can be used to {\em find solutions with probabilistic 
quality guarantees}. 

{\bf Results:} We conduct experiments on two real datasets Sina Weibo and 
Twitter, which are the two most popular microblogs in the world, and compare 
our results with existing state-of-the-art methods. We not only demonstrate our 
approach is computationally efficient, but we also show that random walk based 
sampling can produce higher quality candidate sets than vertex sampling. 
Finally, we apply our method to Sina Weibo 
and show the effectiveness of 
sensors' detection and prediction capability on discovering events.

This paper is organized as follows. In Section II, we introduce notations and 
formulate the problem. The basic framework of the proposed method is 
introduced in Section III along with performance guarantee analysis. 
Experiments are conducted in Section IV. Section V summarizes the related 
works, and Section VI concludes.

\section{II. Problem Statement} \label{sec:formulation}

Let us first introduce some notations 
we use in this paper, and then formally present the problem formulation.

\subsection{Notations and Problem Formulation} 

Let $G(V,E,C)$ denote the OSN under study, where $V$ is the set of nodes, $E$ 
is the set of edges, and $C$ is the set of information cascades. An 
\emph{information cascade} (or \emph{cascade} for short) $c\in C$ is represented by 
a set of participating times $\{t_{cu}:u\in V, c\in C, t_{cu}\geq 0\}$, where 
$t_{cu}$ denotes the time that node $u$ first participates cascade $c$. If $u$ 
never joins $c$ during our observation then $t_{cu}=\infty$. The \emph{size} of a 
cascade is the number of users with a finite participating time, denoted by 
$size(c)$, i.e., $size(c)=|\{u:u\in V\wedge t_{cu}<\infty\}|$. Also, let $t_c$ 
denote the time cascade $c$ begins, i.e., $t_c=\min_ut_{cu}$. 
A summary of these 
notations is shown in Table~\ref{tb:Notations}.

\begin{table}[htb]
    \scriptsize
	\centering
	\caption{Frequently used notations.}
	\label{tb:Notations}
	\begin{tabular}{|l|l|}
	\hline
	{\bf Notation} & {\bf Description} \\
	\hline
	$G(V,E,C)$ & Social network, $V$, $E$ and $C$ are node/edge/cascade sets.\\ 
	$t_{cu}$ & The time user $u$ participates cascade $c$. \\
	$t_c$ & The start time of cascade $c$, $t_c=\min_ut_{cu}$. \\
%	$size(c)$ & Size of cascade $c$, $size(c)=|\{u:u\in V\wedge t_{cu}<\infty\}|$.\\
	$Nb(u)$ & Neighboring nodes of node $u$. \\
	$B, B'$ & Budget of finding sensors and candidates respectively. \\ 
	$K$ & Greedy algorithm stops after $K$ rounds, where $K\leq B$. \\
	$F(\cdot)$ & Reward function we want to optimize. \\
	$\SC,\SC_k$ & Set of candidates or candidates at round $k$. \\
	$S,S_k$ & Set of sensors or sensors obtained after round $k$. \\
	$S_\alpha^*$ & Top $\alpha\%$ of the nodes ordered by reward gain 
	decreasingly.\\
	$OPT$ & The optimal solution of problem \eqref{eq:problem}. \\
	$\delta_s(S)$ & Reward gain of node $s$ with respect to $S$. \\
	$\xi_p(\alpha)$ & Sample size with confidence $p$ and percentile $\alpha$.\\
	\hline
	\end{tabular}
\end{table}

The social sensor placement problem is to select a set of nodes $S\subset V$ as 
social sensors within budget $B$, where $|S|\leq B\ll |V|$,
so as to maximize a \emph{reward 
function} $F(S)$, which is a set function $F:2^V\rightarrow \mathbb{R}_{\geq 0}$ 
and satisfies $F(\emptyset)=0$. The optimal sensor set $OPT$ satisfies
\begin{equation}
	OPT=\argmax_{S\subset V\wedge |S|\leq B}F(S). 
	\label{eq:problem}
\end{equation}

The reward function, in general, is determined by requirements of the problem
under study\cite{Leskovec2007}. In this paper, we want to trade off importance 
(quantified by $size(c)$) against timeliness (quantified by $t_{cu}-t_c$), 
which results in the following,
\begin{equation}
	F(S)=\sum_{c\in C}\frac{size(c)}{1+\min_{u\in S}\{t_{cu}-t_c\}}.
	\label{eq:reward}
\end{equation}
That is, if the sensor set $S$ can \emph{cover} 
as many large size cascades as early as 
possible, the reward will be high. 

\subsection{Submodularity and Greedy Algorithm}
Optimization problem \eqref{eq:problem} is NP-complete\cite{Kempe2003}. However, 
when $F$ is a) {\em nondecreasing}, i.e., if $S\subseteq T\subseteq V$, then 
$F(S)\leq F(T)$, and b) {\em submodular}, i.e., if $S\subseteq T\subseteq V$, then 
$F(S\cup\{s\})-F(S)\geq F(T\cup\{s\})-F(T)$, $\forall s\in V\backslash T$, the 
greedy algorithm can obtain an approximate solution that is at least 
$1-1/e\approx 63\%$ of the optimal\cite{Nemhauser1978}. It is easy to show 
that Eq.~\eqref{eq:reward} possesses these two properties. 

The greedy algorithm can be stated as follows.
It runs for at most $B$ rounds to obtain a set $S$ of size $|S|\leq B$. 
In each round, it finds a node $s\in V \backslash S$ that maximizes the 
\emph{reward gain} $\delta_s(S)\triangleq F(S\cup\{s\})-F(S)$, then $s$ is 
added into $S$ in this round. This process repeats $K$ rounds until $|S|=B$ or 
$\delta_s(S)=0$. The computation complexity of this algorithm is $O(K\cdot 
|V|)$. Note that this greedy algorithm is {\em not} scalable 
for large scale OSNs since graphs of these OSNs usually have large 
number of nodes (or $|V|$ is very large). 

To speed up the greedy algorithm, Leskovec et al\cite{Leskovec2007} use a 
Cost-Effective Lazy Forward (CELF) approach, also known as Accelerated Greedy 
(AG) proposed in \cite{Minoux1978} to reduce the computation times of 
$\delta_s(S)$ in each round by further utilizing the submodularity of 
$F$. The basic idea is that the reward gain of a node in the current round 
cannot be better than its reward gain in the previous round, i.e., if $k>l$, 
then $\delta_s(S_k)\leq \delta_s(S_l),\, \forall s\in V\backslash S_k$, where 
$S_k, 1\leq k\leq K$ is the selected nodes after the $k$-th round by the 
greedy algorithm. However, 
it is important to note that AG/CELF does {\em not} guarantee an improvement on 
computational efficiency, 
i.e., in the worst case, it is as inefficient as the naive greedy 
algorithm\cite{Minoux1978}.

\section{III. Search Space Reduction} \label{sec:methods}

The inefficiency of the above mentioned algorithms is due to the large search
space and the absence of an efficient search method. For example, to find at
most $B$ sensors from the node set $V$, there are 
$\sum_{n=1}^{B}\binom{|V|}{n}=O(2^{|V|})$ possible solutions. Hence, we 
modify our problem, and consider how to find some {\em acceptable good solutions} at a much 
lower computational cost. In the following, we 
first describe the basic framework of our 
approach and present the definition of acceptable good solution of our
algorithm. We then formally show the performance guarantees of the proposed 
approach, along with its variants and cost analysis. 

\subsection{The Basic Framework}
In order to reduce search space, we consider using a candidate set $\SC_k 
\! \subseteq \! V \backslash S_{k-1}$ to represent the search space at round $k$.
This forms the basic search space reduction framework as described in 
Alg.~\ref{alg:fm} (We sometimes suppress the subscript $k$ if there is no 
ambiguity). 

\begin{algorithm}
	\footnotesize
	\label{alg:fm}
	\caption{The basic framework}
	\KwIn{Nodes $V$, budget $B$.}
	\KwOut{Sensors $S$.}
	$S=\emptyset$\;
	\While{$|S|<B$,}{
		Generate candidate set $\SC \subseteq V\backslash S$\; \label{lin:diff}
		Select a node $\hat{s}^*$ such that $\hat{s}^*=\argmax_{s\in 
		\SC}\delta_s(S)$\;
		$S=S\cup \{\hat{s}^*\}$\;
	}
\end{algorithm}

The only difference between Alg.~\ref{alg:fm} and the original greedy algorithm is 
the process at line~\ref{lin:diff}. In the original algorithm, one selects a node 
$s^*$ from $V\backslash S$ that maximizes the reward gain $\delta_s(S)$. Here, we 
select $\hat{s}^*$ from a candidate set $\SC\subseteq V\backslash S$ to maximize 
$\delta_s(S)$. Intuitively, the accuracy of this algorithm should be arbitrary 
close to the greedy algorithm as $\SC \rightarrow V\backslash S$ at each round but 
with a reduction in computational cost. Furthermore, if the sample size at each 
round is the same, then Alg.~\ref{alg:fm} becomes $|V|/|\SC|$ times 
faster than the original greedy algorithm. 
In later sub-section, we will present algorithms on how to generate the candidate set $\SC$.
Let us first define what we mean by acceptable good solutions.

\subsection{Acceptable Good Solutions}
At round $k$, suppose we rank the nodes in $V\backslash S_{k-1}$ by reward gain in 
a decreasing order, and denote the top 
$\alpha\%$ of the nodes by $S_\alpha^*$, 
where $\alpha \in (0,100]$.
If 
$\hat{s}^*$ falls into $S_\alpha^*$, we say $\hat{s}^*$ is an \emph{acceptable good 
sensor} found in round $k$. 
It is obvious that one needs to set $\alpha$ to be small, say less than 1,
so to find acceptable good sensors.
All acceptable good sensors found after $K$ rounds form 
the \emph{acceptable good solution} of Alg.~\ref{alg:fm}.

The candidate set size $|\SC|$ will affect the likelihood that at least one node in 
$\SC$ falls into $S_\alpha^*$. Formally, we can calculate the probability 
that at least $k$ nodes in $\SC$ falls into $S_\alpha^*$ as 
\begin{equation}
 Prob\{|\SC\cap S_\alpha^*|\geq 
 k\}=\sum_{i=k}^{|S_\alpha^*|}\frac{\binom{|S_\alpha^*|}{i}\binom{|V\backslash 
 S_\alpha^*|}{|\SC|-i}}{\binom{|V|}{|\SC|}}, 
 \label{eq:ssk}
\end{equation}
which follows a hypergeometric distribution. For $k=1$, this is equivalent to
\begin{equation}
	Prob\{|\SC\cap S_\alpha^*|\geq 1\}=1-(1-\alpha\%)^{|\SC|}. 
	\label{eq:sample_size_1}
\end{equation}
In order to achieve a confidence level that $Prob\{|\SC\cap S_\alpha^*|\geq 1\}\geq 
p$, we can determine the lower bound on $|\SC|$ as
\begin{equation}
	|\SC|\geq\left\lceil \frac{\ln(1-p)}{\ln(1-\alpha\%)}\right\rceil.
\end{equation}

\begin{figure}
	\centering
	\subfloat[Candidates size $|\SC|$.\label{fig:ss}]
		{\includegraphics[width=0.5\linewidth]{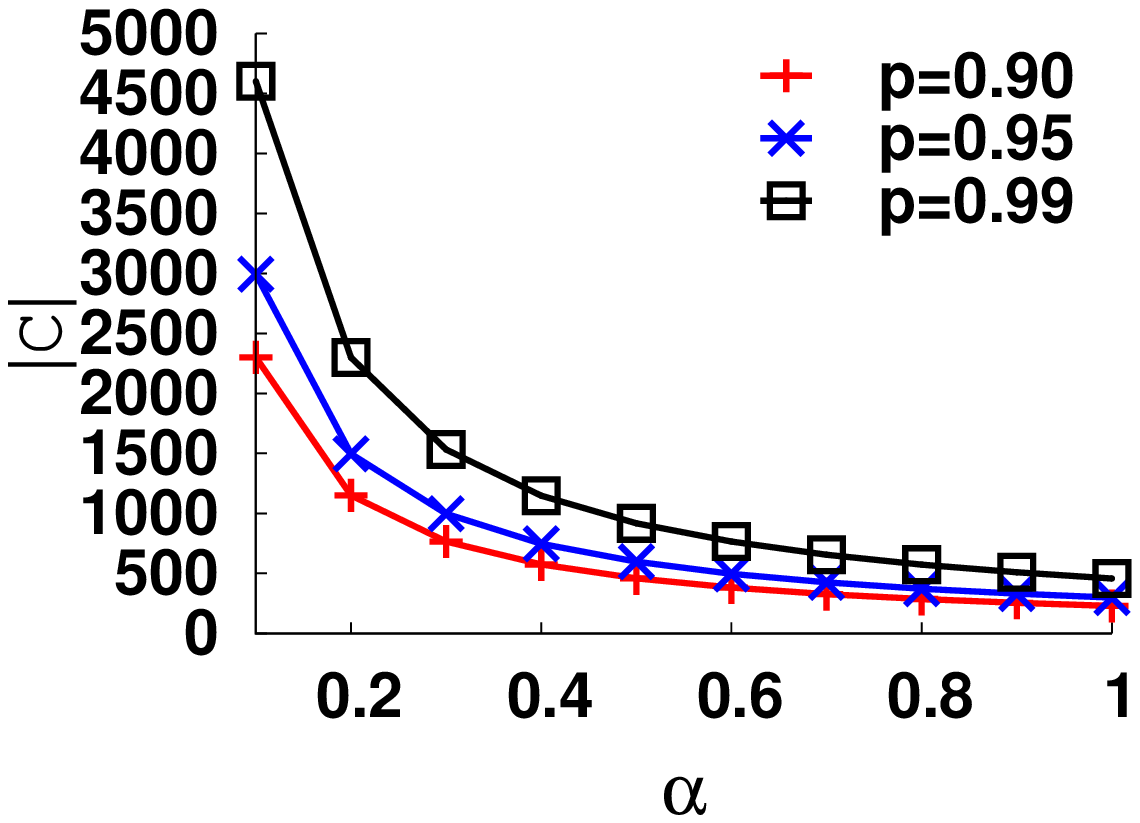}}
	\subfloat[Overlap percentile.\label{fig:oe}]
		{\includegraphics[width=0.5\linewidth]{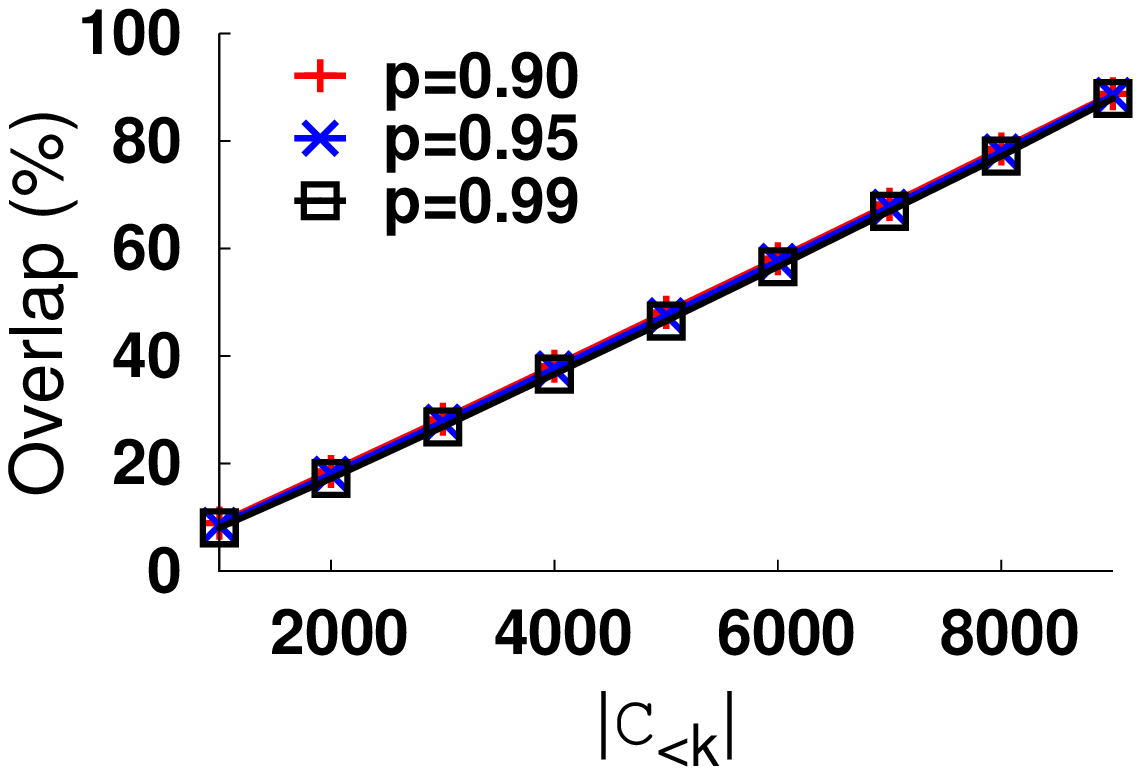}}
	\caption{(a) Reduction of $|\SC|$ as we increase $\alpha$.
         %Sample size and overlap. 
         (b) For a small network of $|V|=10000$ and $|\SC_k|=1000$ nodes,
         there is a high degree of overlap.}
	\label{fig:so}
\end{figure}

Fig.~\ref{fig:ss} shows the relationship between $|\SC|$ and $\alpha$ with three 
different confidence levels of $p$, which we set 
to 0.90, 0.95 and 0.99. One interesting observation is 
that, when $\alpha$ varies from 0.1 to 1
(or top 0.1\% to 1\%), $|\SC|$ drops quickly, which means that 
the search space size can be dramatically reduced. For example, if we want to 
choose a node in top 1\% (or $\alpha\!=\!1$) of the best nodes from the whole 
population, then one can be sure that one of the 458 nodes will be a good sensor 
with probability greater than 0.99. There is one technical issue 
we need to pay attention:

\noindent\textbf{Dealing with small networks.} 
When the network is small, the advantage of 
using our framework is not significant. We can 
further improve the efficiency of Alg.~\ref{alg:fm} by exploiting the submodular 
property of reward functions as used in \cite{Minoux1978,Leskovec2007}. One 
conclusion from Eq.~\eqref{eq:ssk} is that, when $|V|$ is small, at round $k$, the 
overlap between previous candidates $\SC_{<k}=\bigcup_{i=1}^{k-1}\SC_i$ and current 
candidates $\SC_k$, i.e., $|\SC_{<k}\cap\SC_k|$, is large and increases as 
$|\SC_{<k}|$ increasing. An example is illustrated in Fig.~\ref{fig:oe}. 
In the example, we have a small network
and its size is set to $|V|=10000$. At each round, $1000$ candidates 
are selected. 
We show the least overlap percentile  versus selected candidate set 
size $|\SC_{<k}|$ with a given confidence $p$. 
For example, when $1000$ candidates have 
been selected in the previous rounds, 
for a newly $1000$ selected candidates, there 
is a guarantee that at least 
8\% of them have been selected in previous rounds with 
probability greater than 0.99. For those overlapped nodes, we can reduce the 
calculations of updating reward gain by utilizing the submodular property. 
This will reduce the computational complexity of our framework.

To achieve this, we can store the candidate node $u$ and its reward gain $\delta_u$ 
in a tuple $\langle u,\delta_u,\#_u\rangle$. Here $\#_u$ is the round during which 
the reward gain of node $u$ is calculated or updated. For candidates $\SC_k$, we 
only need to calculate the reward gain of {\em newly selected nodes} and we do not 
need to update the reward gain of previously selected nodes \emph{immediately}. We 
arrange nodes in $\SC_k$ as a priority queue $Q$ by descending order of reward 
gain. Then we access the head $v$ of $Q$ to check whether $\#_v$ equals to the 
current round $k$. If yes, then $v$ is added into $S$; otherwise, update $v$'s 
reward gain, and put $v$ into $Q$ again. This way, the times of updating reward 
gain of sampled nodes can be reduced and Alg.~\ref{alg:fm} will be at least as 
efficient as AG/CELF when dealing with small networks.

\begin{comment}
\noindent\textbf{Issue 2: Terminating criterion.}
{\color{red} The original greedy algorithm can terminate before $B$ sensors have 
been selected. It can be proved that the greedy algorithm finds the optimal 
solution in this case\cite{Nemhauser1978}. This is not the case with 
Alg.~\ref{alg:fm}, it is possible that reward gains of all the sampled nodes are 0, 
but one cannot guarantee that an optimal solution is found in this case. This issue 
can be avoided by carefully implementing the sampling method, e.g., make sure that 
the newly sampled nodes contains new cascades; otherwise, the algorithm terminates 
after several tries.}{\color{blue} This part is not clear. I need to do some 
calculations here.}
\end{comment}

\subsection{Performance Analysis}
We conclude from our previous discussion that we can probabilistically guarantee 
in selecting an acceptable sensor at each round from a small candidate set $\SC$. 
Here, we quantify the quality of the final sensor set obtained after $K$ rounds. 
That is, {\em how close the solution obtained by Alg.~\ref{alg:fm} is to the 
optimal solution of problem \eqref{eq:problem}}? The following theorem answers this 
question. 

\begin{theorem}\label{th:bounds}
Denote the set of sensors obtained by Alg.~\ref{alg:fm} after round $k$ as 
$S_k=\{\hat{s}_1^*,\cdots,\hat{s}_k^*\}$, $1\leq k\leq K$. Let 
$\lambda_k=\delta_{\hat{s}_k^*}(S_{k-1})/\delta_{s_k^*}(S_{k-1})\in 
(0,1]$, where $s_k^*=\argmax_{s\in V\backslash S_{k-1}}\delta_s(S_{k-1})$.
Let $\lambda=\min_{1\leq k\leq K}\lambda_k$, then 
\begin{align}
	F(S_K)\geq (1-\frac{1}{e^\lambda})F(OPT).
\end{align}
\end{theorem}

\noindent
{\bf Proof:}
Please refer to the Appendix.
\done

\noindent
{\bf Remark:}
Theorem~\ref{th:bounds} indicates an important property of 
Alg.~\ref{alg:fm}, 
i.e., if at each step the reward gain $\delta_{\hat{s}^*}(S)$ is bounded by a 
factor $\lambda$, then the final solution is bounded by another factor 
$1-1/e^\lambda$. 
When $\lambda \approx 1$, our final solution is
$1\!-\!1/e \approx 63\%$ of the $OPT$.
Hence, Alg.~\ref{alg:fm} guarantees a good solution when $\lambda$ 
is close to one. Note that in general, one cannot guarantee that $\lambda$ is close 
to one all the time. For example, the reward gain of the second best node is much 
less than the best node in current round. In such a case, finding the best node is 
like finding a needle in a haystack, which illustrates the {\em intrinsic 
difficulty} due to the reward distribution. Nevertheless, one can still accept the 
approximate solution if we believe the second best solution 
has a reward which is not significantly different from the best solution.
%{\em good enough}.

%We analyze the quality of Alg.~\ref{alg:fm} from another point of view. 
One can also understand the quality of Alg.~\ref{alg:fm} from another point
of view.
We define 
{\em cover ratio} as the fraction of nodes in $OPT$ that are in 
all candidate sets, 
i.e., $r=|\bigcup_{k=1}^K\SC_k\cap OPT|/K$. Intuitively, the higher the cover 
ratio, the better the quality of the final solution. 
Assume the size of candidate set 
equal in each round, and denote it by $\xi_p(\alpha)$ with confidence $p$ and 
percentile $\alpha$. The following theorem presents a lower bound on the 
expectation of cover ratio for fixed $p$ and $\alpha$. 

\begin{theorem}\label{th:overlap}
If we set $\alpha=100K/|V|$ at each round, then $\E{r}\geq 1-1/e^p$.
\end{theorem}
\noindent
{\bf Proof:}
Please refer to the Appendix.
\done

\noindent
{\bf Remark:}
Because $p$ can be very close to 1, $1-1/e^p$ will be very close to $1-1/e\approx 
0.63$. In fact, the cover ratio can be further improved by increasing sample size 
at each round, e.g., we can derive in a similar manner that if we set 
$\alpha=50K/|V|$, then the expected cover ratio will be at least 0.82,
or 82\% of those sensors in $OPT$.

\subsection{Generating $\SC$ via Graph Sampling Methods}

%In previous discussion, $\SC$ is generated by blindly picking and doesn't exploit 
%any structural information of OSNs. 
In previous discussion, we did not specify precisely how to generate 
the candidate set $\SC$.  In fact, $\SC$ can be generated by
randomly selecting nodes in $V$.  This does not exploit
any structural properties of OSNs, 
hence the performance discussed in previous sub-section 
can be viewed as the {\em worst case guarantee}.
%should be the worst case. 
Here, we show how to use {\em graph sampling} to generate 
$\SC$. The main advantage of using graph sampling is that one does not need 
to know the complete graph topology prior to executing the sensor placement 
algorithm. This is one main advantage of our framework as compare with
the current state-of-the-art approaches\cite{Minoux1978,Leskovec2007}.
Furthermore,
we also want to explore how different graph sampling methods may affect 
sensor quality. Hence, we make two 
modification to Alg.~\ref{alg:fm}: 
\begin{itemize}
\item Instead of selecting new candidates in each round, 
      we construct one candidate set $\SC$ at the beginning of the algorithm; 
%\item After including a node $v$ in $\SC$, we select the next node $s$ 
%       from $v$'s neighbors (or $Nb(v)$) which can maximize the reward gain 
%       and we put $s$ into $\SC$. 
\item When a node $v$ is sampled, we select a candidate node, say $s$, from $v$ or one of $v$'s
neighbors (i.e., $\{v\} \cup Nb(v)$) which can maximize the reward gain and we put $s$ into 
$\SC$. 
\end{itemize}
The second change is useful in filtering out noisy tweets in OSNs (e.g., tweets 
containing unimportant cascades), and reduces the candidate set size. 
Nodes with 
large reward gain in the neighborhood are more preferred to be in $\SC$. 

Vertex sampling (VS) and random walk (RW) are two popular graph sampling methods. 
We design two variants of Alg.~\ref{alg:fm} based on vertex sampling and random 
walk respectively, they are illustrated in Algs.~\ref{alg:vs} and \ref{alg:rw}. 

\begin{algorithm}[htb]
	\label{alg:vs}
	\footnotesize
	\caption{Combining with vertex sampling}
	\KwIn{Network $G$, sensor budget $B$, candidate budget $B'$.}
	\KwOut{Sensors $S$.}
	$\SC=\emptyset$\;
	\While{$|\SC|<B'$,}{
		Choose a node $v\in V\backslash \SC$\;\label{l:vs}
		Select a node $s^*$ such that $s^*=\argmax_{s\in 
		Nb(v)\cup\{v\}}\delta_s(\SC)$\;
		$\SC=\SC\cup \{s^*\}$\;
	}
	$S=Greedy(\SC,B)$ \tcc*{choose $B$ sensors from $\SC$ using the greedy algorithm}
\end{algorithm}
\begin{algorithm}[htb]
	\label{alg:rw}
	\footnotesize
	\caption{Combining with random walk}
	\KwIn{Network $G$, sensor budget $B$, candidate budget $B'$.}
	\KwOut{Sensors $S$.}
	$\SC=\emptyset$\;
	Set $u=u_0$ which is randomly chosen from $V$\;
	\While{$|\SC|<B'$,}{
		Choose a node $v\in Nb(u)$\;\label{l:rw}
		Select the node $s^*$ such that $s^*=\argmax_{s\in 
		Nb(v)}\delta_s(\SC)$\;
		$\SC=\SC\cup\{s^*\}$\;
		Set $u=v$\;
	}
	$S=Greedy(\SC,B)$ \tcc*{choose $B$ sensors from $\SC$ using the greedy algorithm}
\end{algorithm}

Each of these algorithms contains two steps. In the first step, the candidate set 
$\SC$ with budget $B'$ are constructed. In the second step, the final sensors $S$ 
are chosen. In line~\ref{l:vs} of Alg.~\ref{alg:vs} and line~\ref{l:rw} of 
Alg.~\ref{alg:rw}, we can use various attribute information within an OSN to {\em 
bias} the selection. Such attributes of a user can be the number of 
posts/friends/followers and so on. For example, when using random walk to build the 
candidate set (Alg.~\ref{alg:rw}), we can choose a neighboring node with 
probability proportion to its degree or activity (\#posts), which will bias a 
random walk toward high degree or activity nodes. Intuitively, large degree nodes 
are more likely to be information sources or information hubs, and high activity 
nodes are more likely to retweet tweets. The comparison of using different 
attributes to bias our node selection will be discussed in Section IV. Although 
vertex sampling cannot be biased without knowing attributes of every node in 
advance, in order to study functions of different attributes, we will assume we 
know this complete information and let vertex sampling be biased by different 
attributes. 

\subsection{Sampling Cost Analysis}
The computational cost of social sensor selection is defined to be the number of 
times that the reward gain of nodes is calculated or updated. Because of the second 
change in Alg.~\ref{alg:vs} and Alg.~\ref{alg:rw}, there will be additional cost in 
obtaining samples $\SC$. Here, we analyze the cost of sampling using uniform vertex 
sampling (UVS) (for Alg.~\ref{alg:vs}) and uniform random walk (URW) (for 
Alg.~\ref{alg:rw}). 

For UVS in Alg.~\ref{alg:vs}, the average cost to obtain candidate set $\SC$ is 
\[
	Cost_\text{UVS}=\E{\sum_{i=1}^{B'}(d_i+1)}=B'(1+\E{d_\text{UVS}}), 
\]
where $d_i$ is the degree of $i$-th node in $\SC$, and $\E{d_\text{UVS}}$ is the 
average node degree obtained by UVS. Let $M$ denote the maximum degree 
in the network, and $\theta_d$ the fraction of nodes with degree $d$. Then 
$\E{d_\text{UVS}}=\sum_{d=1}^M d\theta_d\triangleq d_\text{avg}$. 
For power law networks with degree distribution $\theta_d=\frac{1}{Z}d^{-a}$, we 
obtain 
\begin{equation}
	\E{d_\text{UVS}}=\frac{1}{Z}\sum_{d=1}^M d^{1-a}.
	\label{eq:uvs}
\end{equation}

For URW, the average cost to obtain candidate set $\SC$ can be derived 
similarly, that is 
\[
	Cost_\text{URW}=\E{\sum_{i=1}^{B'}d_i}=B'\cdot\E{d_\text{URW}}, 
\]
where $\E{d_\text{URW}}$ is the average node degree under URW, i.e., 
$\E{d_\text{URW}}=\sum_{d=1}^M d\frac{d\theta_d}{d_\text{avg}}$. For power law 
networks, it becomes
\begin{equation}
	\E{d_\text{URW}}=\frac{1}{Zd_\text{avg}}\sum_{d=1}^M d^{2-a}.
	\label{eq:urw}
\end{equation}

Equations \eqref{eq:uvs} and \eqref{eq:urw} reveal the difference between UVS 
and URW. Both of them are functions of $a$, the exponent of the power law degree 
distribution. The difference is that, for UVS, $\E{d_\text{UVS}}$ is always 
finite when $a>2$, which means that the average cost of obtaining a sample 
equal the average degree, and it is usually small and bounded by the Dunbar 
number. For URW, $\E{d_\text{URW}}$ is finite if $a>3$; otherwise the average 
cost can become arbitrarily large as $M$ increases. 

To limit the cost of creating candidate set, we can use simple heuristics to avoid 
searching a local maximum reward gain node from all the neighbors. For example, we 
can limit the searching scope to the top $n$ most active neighbors. In our 
experiments, $n$ can be set very small (e.g., 10), so the cost of obtaining samples 
will be bounded by $O(B'\cdot n)$.

\section{IV. Experiments} \label{sec:experiment}

To evaluate the performance of previous discussed methods, 
we present experimental 
results on two datasets collected from Sina Weibo and Twitter, 
respectively. Then, 
we conduct experiments on Sina Weibo to study the 
{\em detection capability}, which measures
the ability to detect cascades,
as well as {\em prediction capability},
which measures the ability to capture future cascades
using the senors selected based on historial data. 

\begin{table}
	\centering
	\caption{Dataset summary}\label{tab:dataset}
	\begin{tabular}{|l|r|r|}
		\hline
		Dataset & Sina Weibo & Twitter \\ 
		\hline
		Nodes & 0.3M & 1.7M \\
		Edges & 1.7M & 22M \\
		Cascades & 11M & 7M \\
		\hline
	\end{tabular}
\end{table}

\subsection{Experiment on Sina Weibo and Twitter}

\subsubsection{Dataset}
Sina Weibo is one of the most popular microblogging sites in China. Similar to 
Twitter, users in Weibo are connected by the \emph{following} relationships. 
Tweets can be retweeted by one's followers and form cascades. We collected a 
portion of Weibo network using the Breath First Search (BFS) method along the 
following relationships. For a user, 
his tweets and neighbors are all collected. We extract URL links 
contained in tweets, and consider them as the representation of 
cascades. The Twitter dataset contains tweets and network, which are from 
\cite{TwitterDataset} and \cite{TwitterNetwork}, respectively. Similar 
to Weibo, 
URL links and hashtags contained in tweets are extracted to form cascades. 
Table~\ref{tab:dataset} summarizes the statistics of these two networks.
%The basic statistics of the two networks are summarized in Table~\ref{tab:dataset}. 

\subsubsection{Evaluating the basic framework}
We first evaluate the performance of Alg.~\ref{alg:fm} without any network
information. We choose $B$ sensors where $B$ ranges from 0.1\% to 1\% of the total
number of nodes, and compare the quality of sensors and speed-up of the algorithm
with the greedy algorithm and AG/CELF. Speed-up of an algorithm is defined as 
$\frac{Cost(\text{greedy algorithm})}{Cost(\text{algorithm})}$, where 
$Cost(\text{A})$ represents the cost of algorithm A, i.e., number of times of 
calculating or updating reward gains. The sample size at each round is fixed to 
$\xi_{0.9}(\alpha)$ and $\xi_{0.9}(0.5\alpha)$ respectively, where 
$\alpha=100B/|V|$.

\begin{figure}[t]
	\subfloat[Reward (Weibo)]{\includegraphics[width=0.5\linewidth]{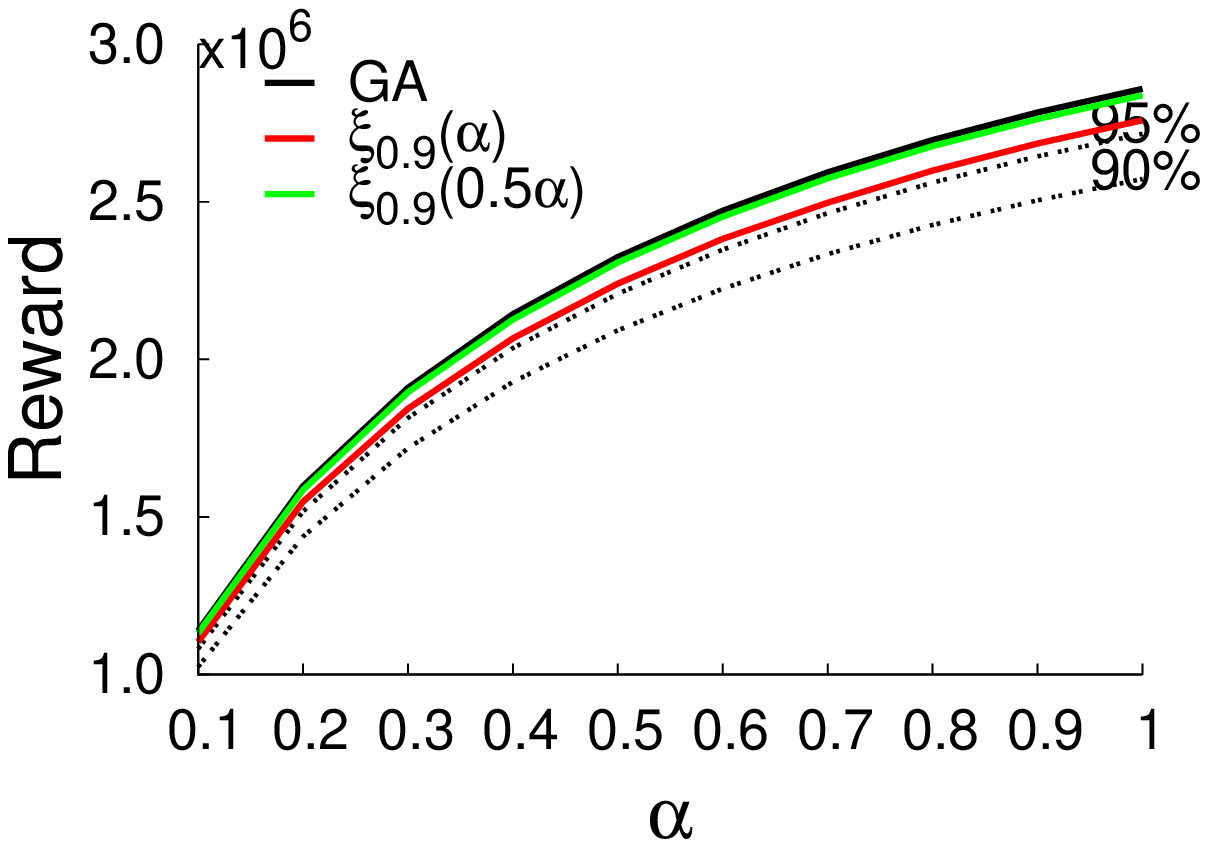}}
	\subfloat[Speed-up (Weibo)]{\includegraphics[width=0.5\linewidth]{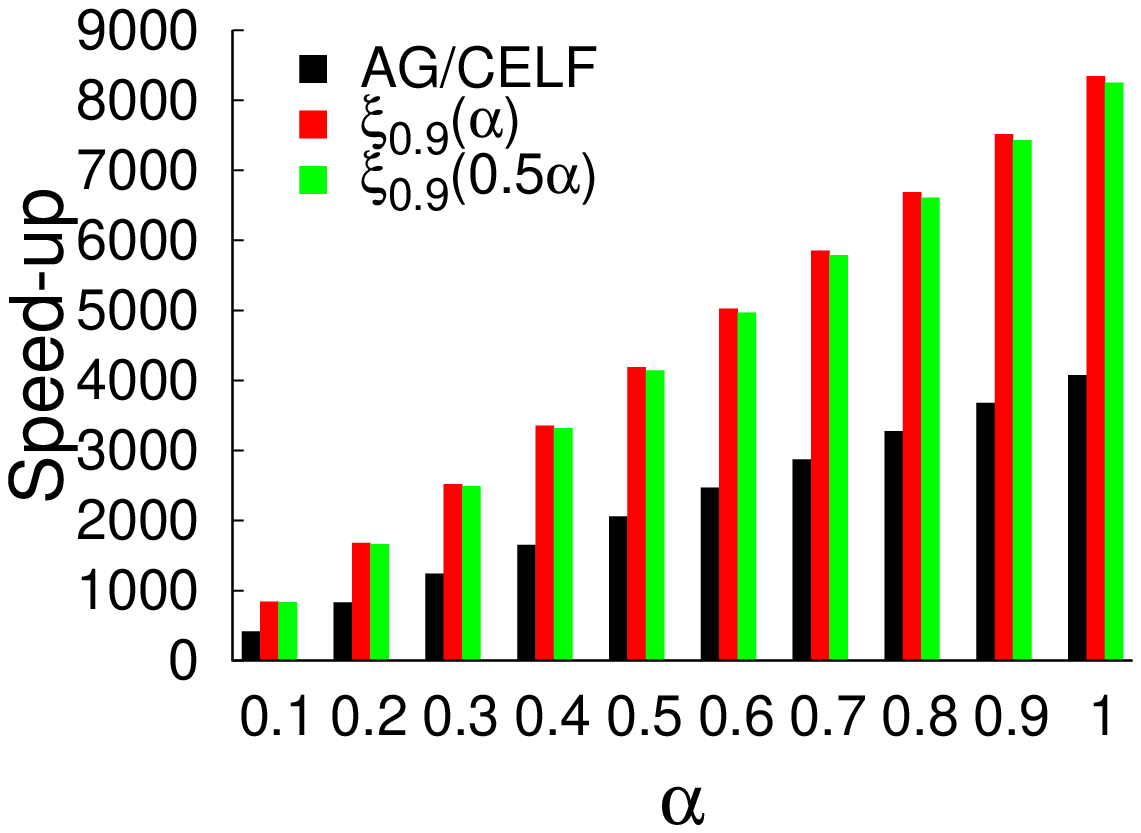}}\\
	\subfloat[Reward (Twitter)]{\includegraphics[width=0.5\linewidth]{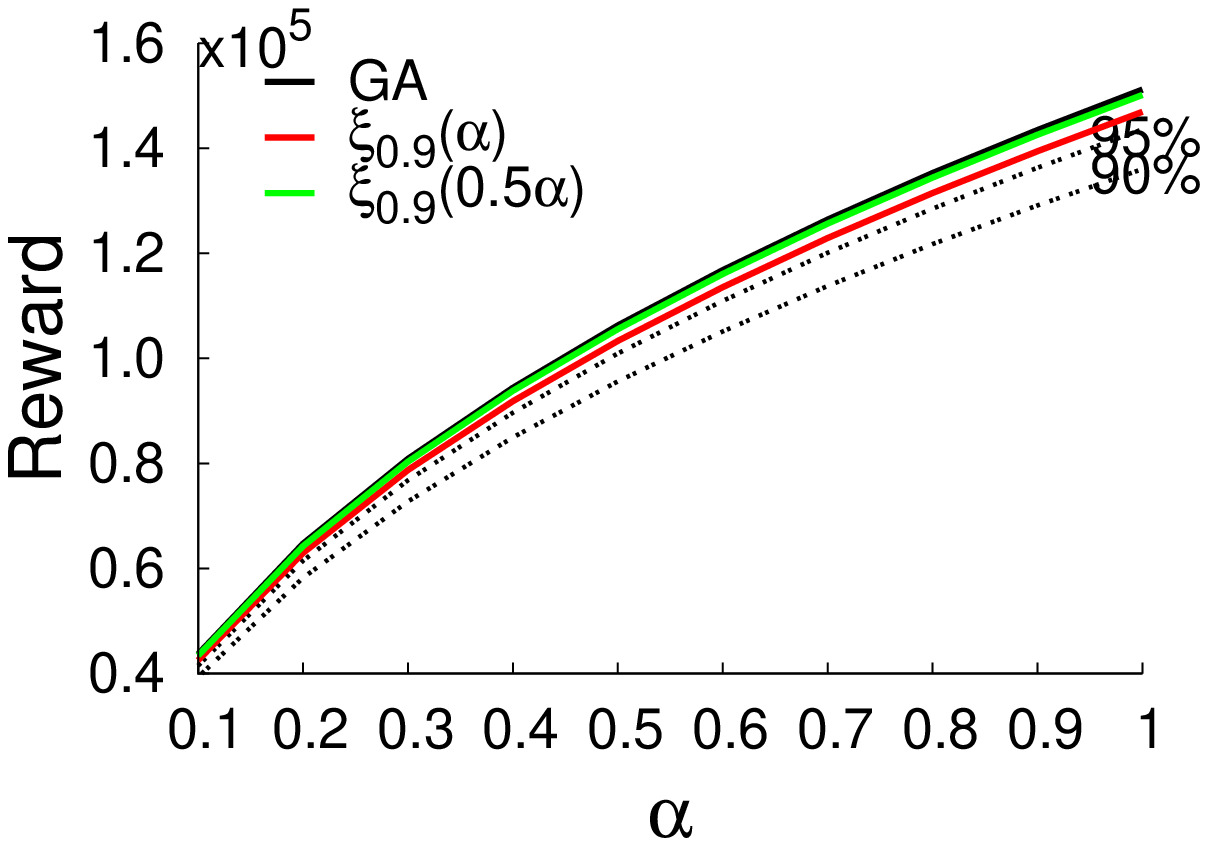}}
	\subfloat[Speed-up (Twitter)]{\includegraphics[width=0.5\linewidth]{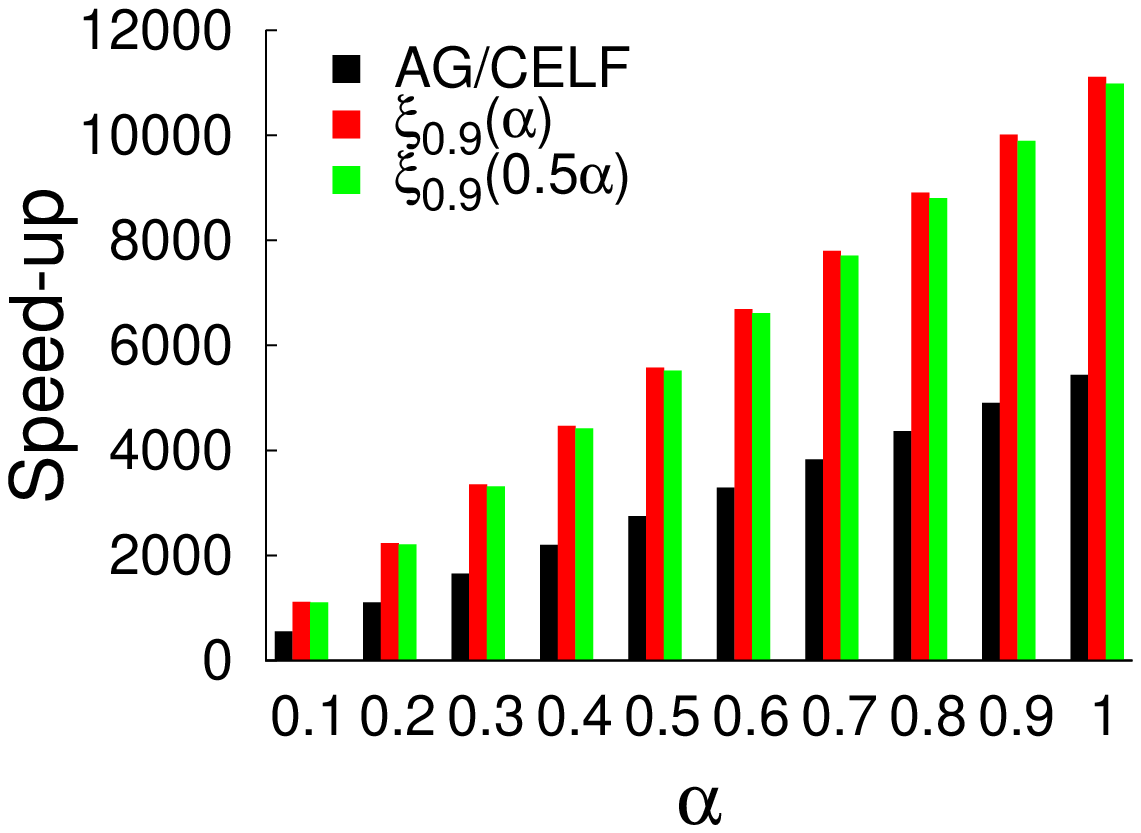}}
	\caption{Evaluating the basic framework on real datasets.}
	\label{fig:bf}
\end{figure}

In the reward curves of Fig.~\ref{fig:bf}, two dashed lines represent 
90\% and 95\% of the total reward by the greedy algorithm respectively. 
We show the 
rewards of sensors with different sizes. 
One can observe that the accuracy in total reward of the 
sampling approach is within 90\% of the greedy algorithm, 
and that it is more computational efficient 
than AG/CELF from the speed curves. When sample size increases from 
$\xi_{0.9}(\alpha)$ to $\xi_{0.9}(0.5\alpha)$, 
the reward increases to about 95\% 
of that produced by the greedy algorithm, 
%but at a higher computational cost.  
but with a slight reduction in speedup.
Hence, one can 
adjust the sample size to trade off between accuracy and efficiency. 

\subsubsection{Evaluating the vertex sampling framework} 
\begin{figure}[t]
	\subfloat[Reward (Weibo)]{\includegraphics[width=0.5\linewidth]{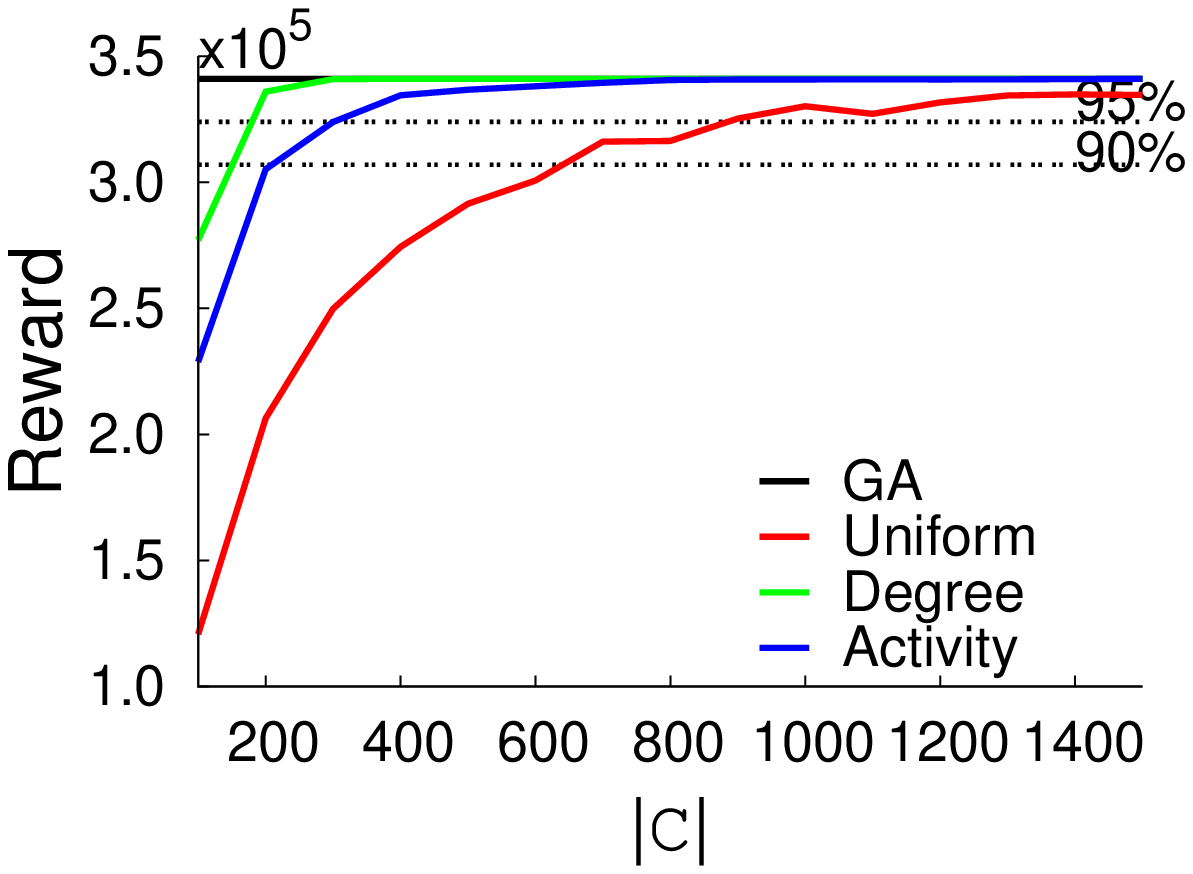}}
	\subfloat[Speed-up (Weibo)]
		{\includegraphics[width=0.5\linewidth]{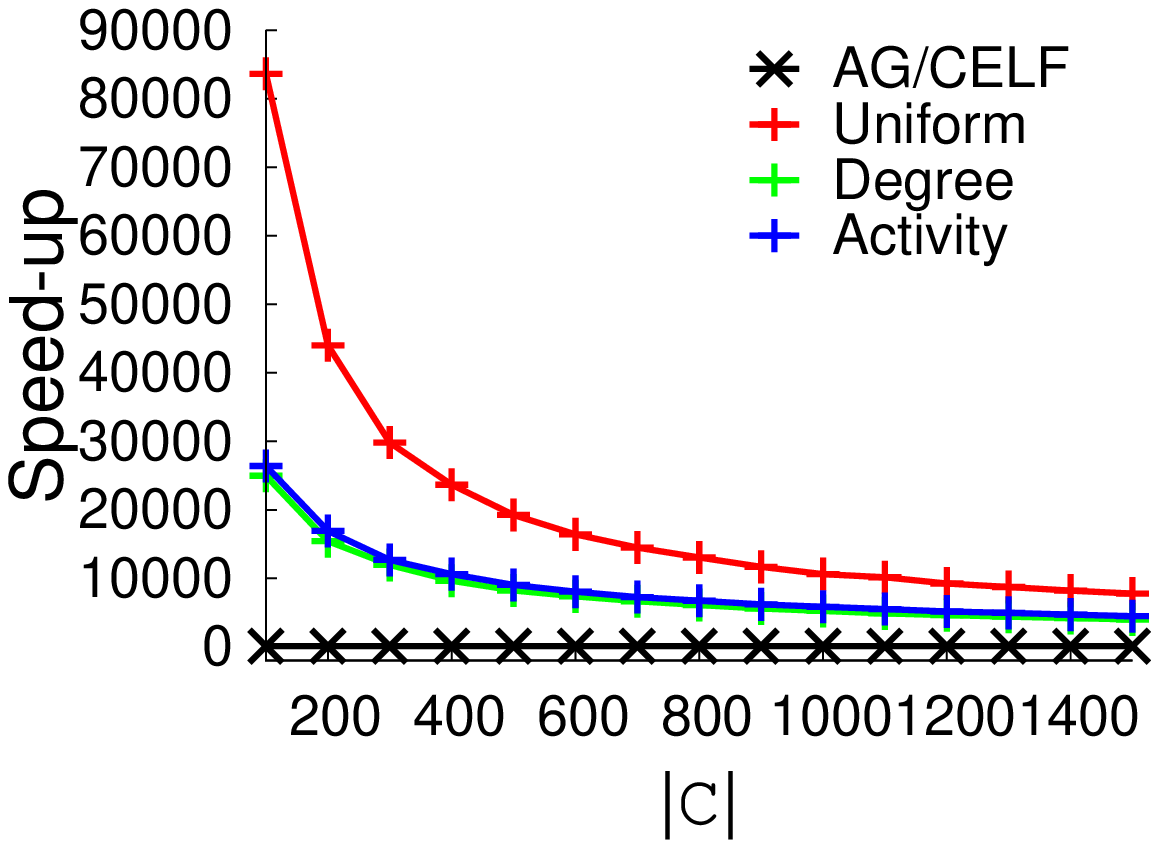}}\\
	\subfloat[Reward (Twitter)]
		{\includegraphics[width=0.5\linewidth]{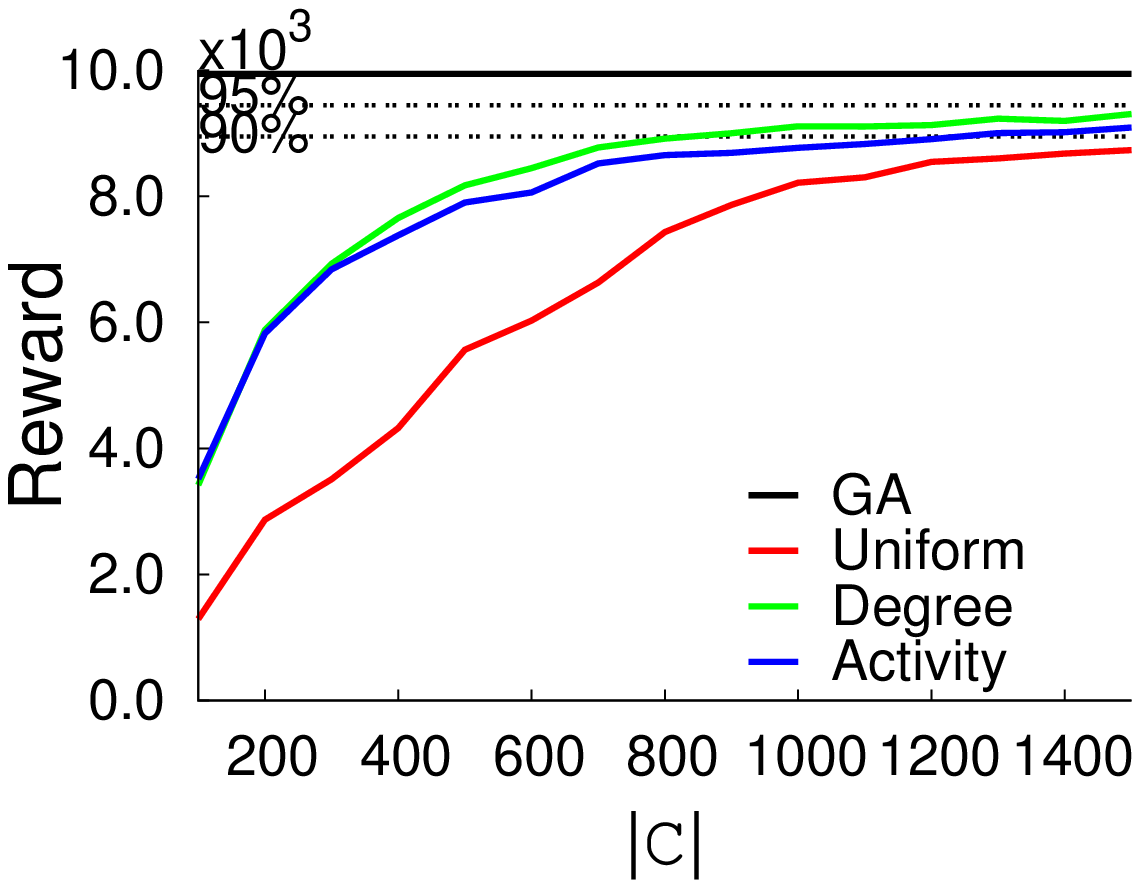}}
	\subfloat[Speed-up (Twitter)]
		{\includegraphics[width=0.5\linewidth]{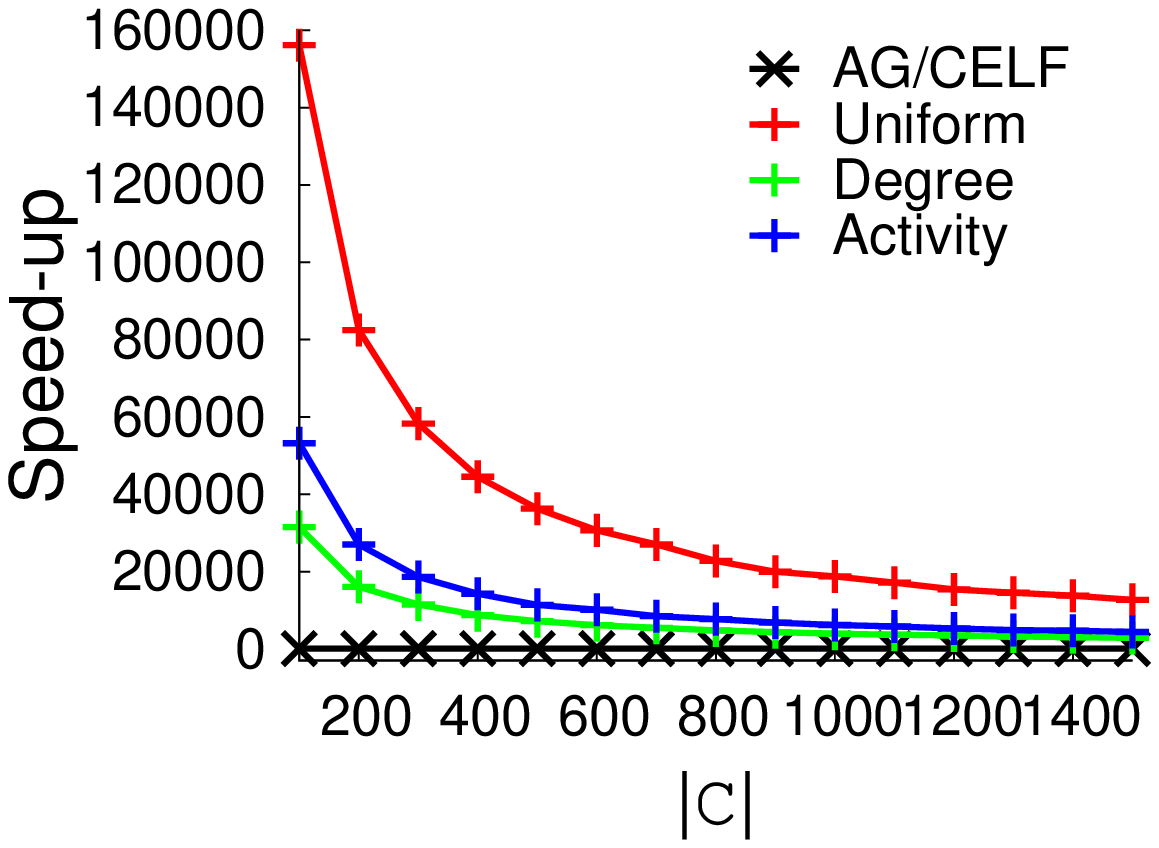}}
	\caption{Vertex sampling on real datasets ($B=100$).}
	\label{fig:vs}
\end{figure}

Next, we evaluate the benefit of using attribute information within an OSN,
in particular, in reducing the size of the candidate set
and reducing the computational cost of Alg.~\ref{alg:vs}.
%Next, we evaluate the benefit of to what extent 
%network information can reduce the 
%sample size and improve efficiency by Alg.~\ref{alg:vs}. 
%The attributes we used
%for biasing vertex sampling include, 
For vertex sampling, we consider the following variants:
(a) uniformly selecting a node from the network;
(b) select a node from the network with a probability proportional to
    its degree;
(c) select the node from the network with a probability proportional to
    its activity (say \# posts).
%uniformly, by degree and by activity(\#posts). 
The aim is to select $B=100$ sensors. Each experiment is run 10 times, and the 
averaged results are shown in Fig.~\ref{fig:vs}.  From the reward curves, we observe 
that vertex sampling by degree is the best follows by sampling by activity, and 
uniform vertex sampling is the worst. However, from the speed-up curves we can see 
that uniform vertex sampling is the most efficient approach. Sampling by degree is 
the most expensive method. But the sampling approaches are in general much more 
efficient than AG/CELF. 

\subsubsection{Evaluating the random walk framework}
\begin{figure}[t]
	\subfloat[Reward (Weibo)]
		{\includegraphics[width=0.5\linewidth]{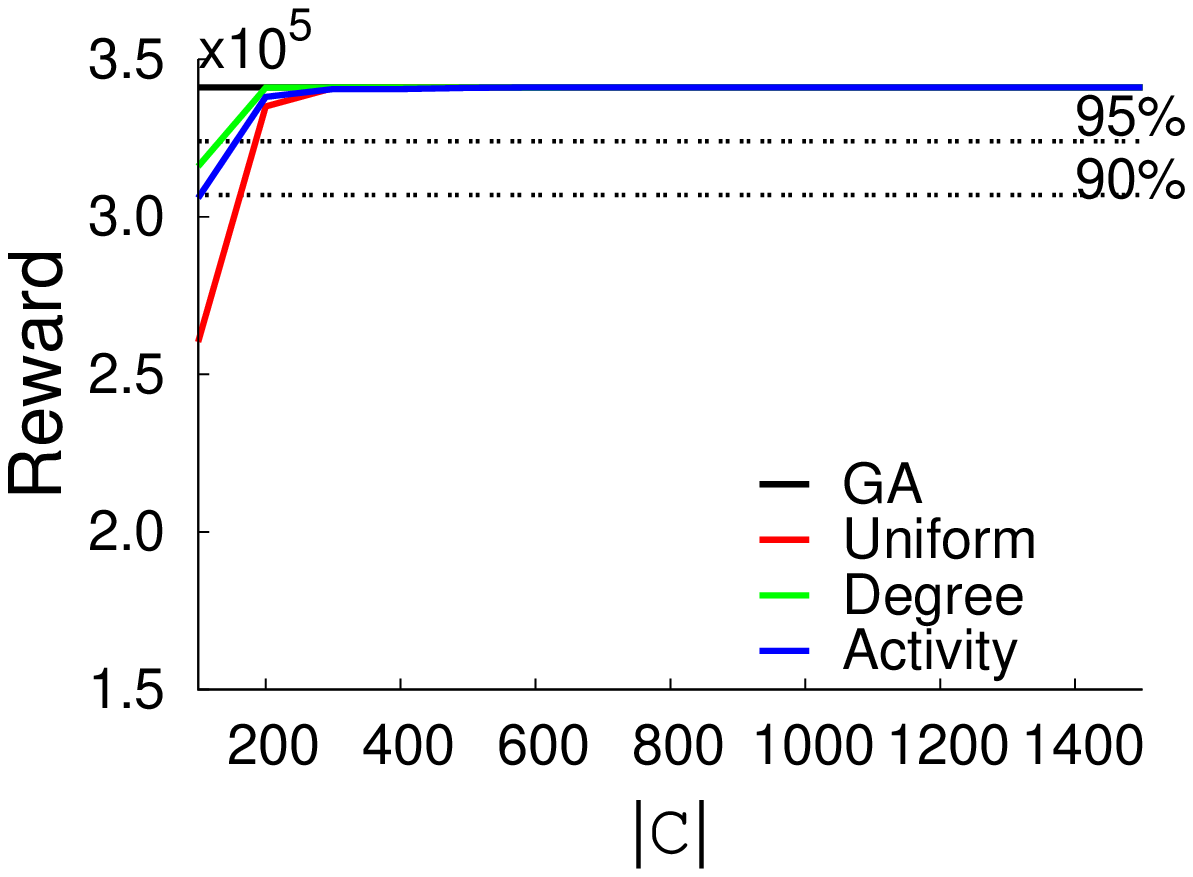}}
	\subfloat[Speed-up (Weibo)]
		{\includegraphics[width=0.5\linewidth]{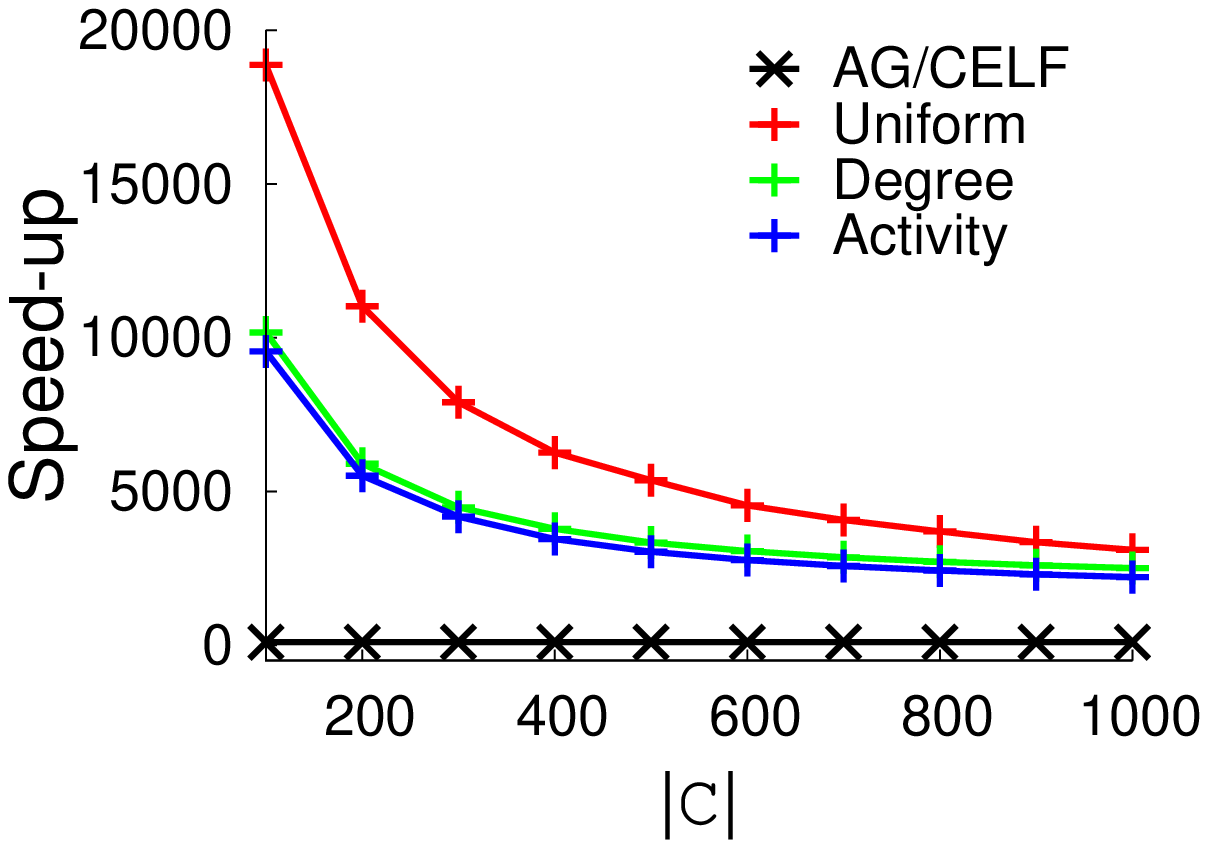}}\\
	\subfloat[Reward (Twitter)]
		{\includegraphics[width=0.5\linewidth]{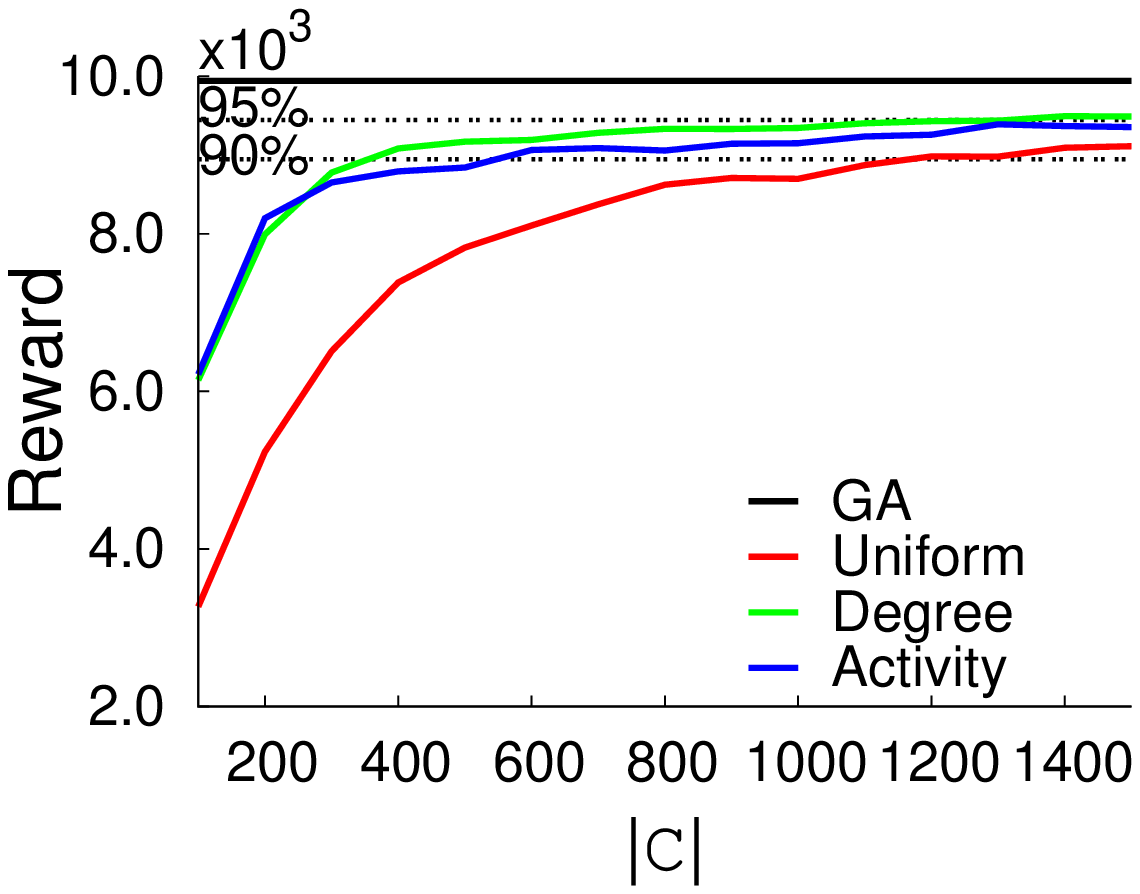}}
	\subfloat[Speed-up (Twitter)]
		{\includegraphics[width=0.5\linewidth]{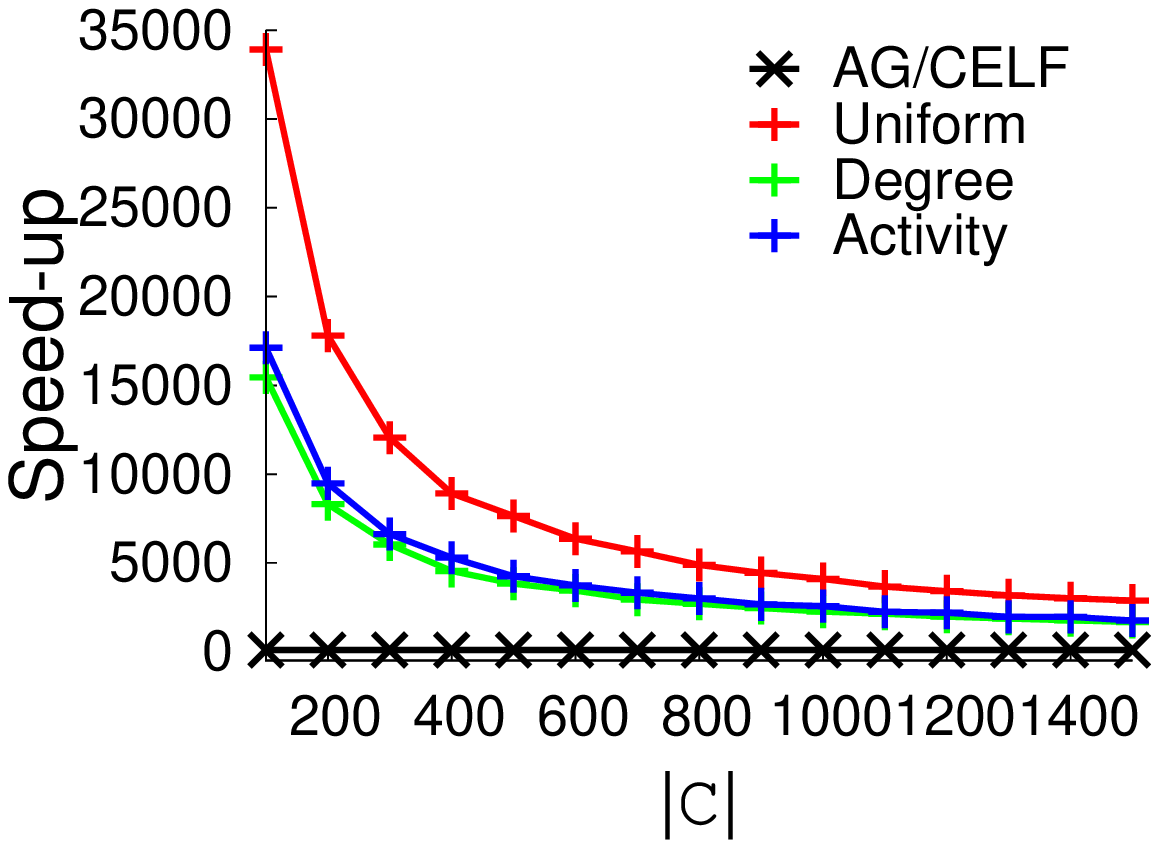}}
	\caption{Random walk on real datasets ($B=100$).}
	\label{fig:rw}
\end{figure}

\begin{figure}
	\subfloat[Weibo]{\includegraphics[width=0.5\linewidth]{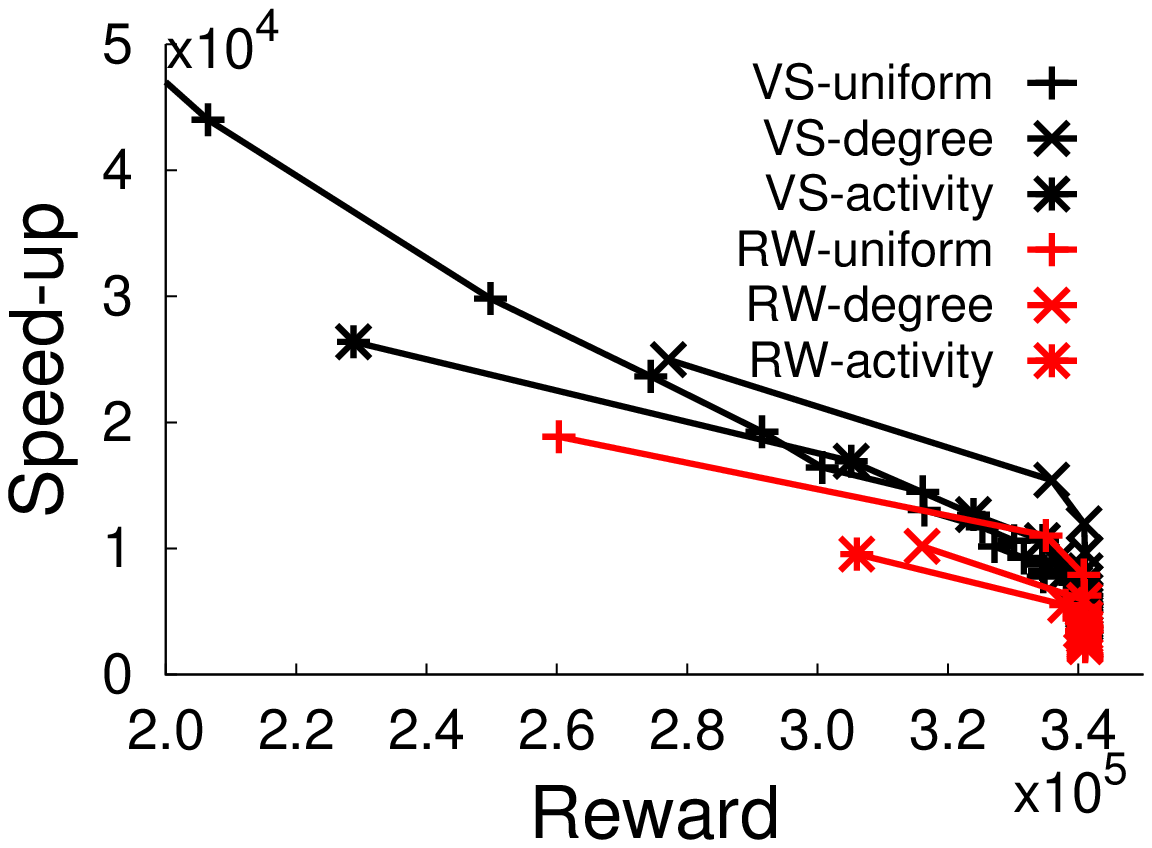}}
	\subfloat[Twitter]{\includegraphics[width=0.5\linewidth]{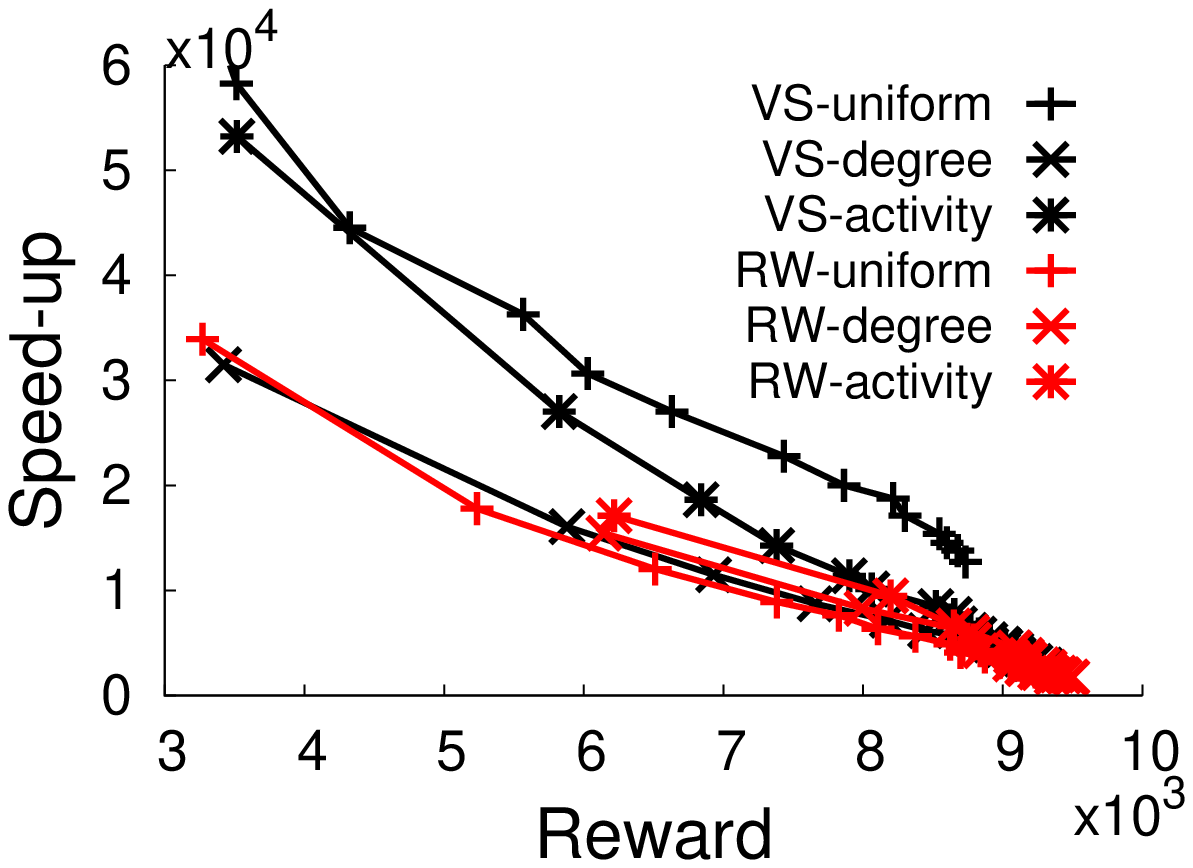}}
	\caption{Comparing vertex sampling with random walk in Speedup and Reward.}
	\label{fig:vsrw}
\end{figure}

We now evaluate the performance of the random walk algorithm
as described in Alg.~\ref{alg:rw}.  
Note that this algorithm does not require the knowledge of the network topology in advance.
We have three variants:
(a) select a neighboring node uniformly;
(b) select a neighboring node with a probability proportional to its degree;
(c) select a neigbhoring node with a probability proportional to its activity.
%biased by uniform/degree/activity 
%described in Alg.~\ref{alg:rw}. 
The other settings are similar to vertex sampling, 
and the results are shown in Fig.~\ref{fig:rw}. Again, we can see that degree is 
the best attribute for random walk, it is also the most expensive one. 
Comparing random walk with vertex sampling, 
we observe that random walk can achieve higher accuracy but at a higher 
computational cost, as is shown in Fig.~\ref{fig:vsrw}.
Both of them are more efficient than AG/CELF. Furthermore, the random walk algorithm
{\em does not} require a full topology beforehand.
%we observe that random walk performs better than 
%vertex sampling, but vertex sampling is more efficient than random walk. Both of 
%them are significantly more efficient than AG/CELF. 

\subsection{Application Study on Sina Weibo}

For a set of selected sensors, we are interested in its detection and prediction 
capabilities. For detection, it means how many or how timely a set of sensors can 
discover information cascades from a given dataset. For prediction, it means how well 
a set of sensors selected based on historical dataset can generalize to a future 
events, e.g., a set of sensors are selected based on a past week's data, and we 
want to know how well they perform on a future week's data. We
conduct experiments on Sina Weibo to study these two capabilities of sensors 
selected by different methods. 

\subsubsection{\textbf{Detection Capability Analysis}}
We compare the detection quality of our algorithm with two baselines: a) randomly choose a 
set of nodes as sensors; b) choosing sensors by utilizing {\em friendship 
paradox}\cite{Christakis2010}. Friendship paradox randomly selects a neighbor of a 
randomly sampled node as a sensor, and it is proved to be able to sample larger 
degree nodes than random node selection\cite{Feld1991}. 

\begin{figure}[t]
	\centering
	\subfloat[\#detects]{\includegraphics[width=0.5\linewidth]{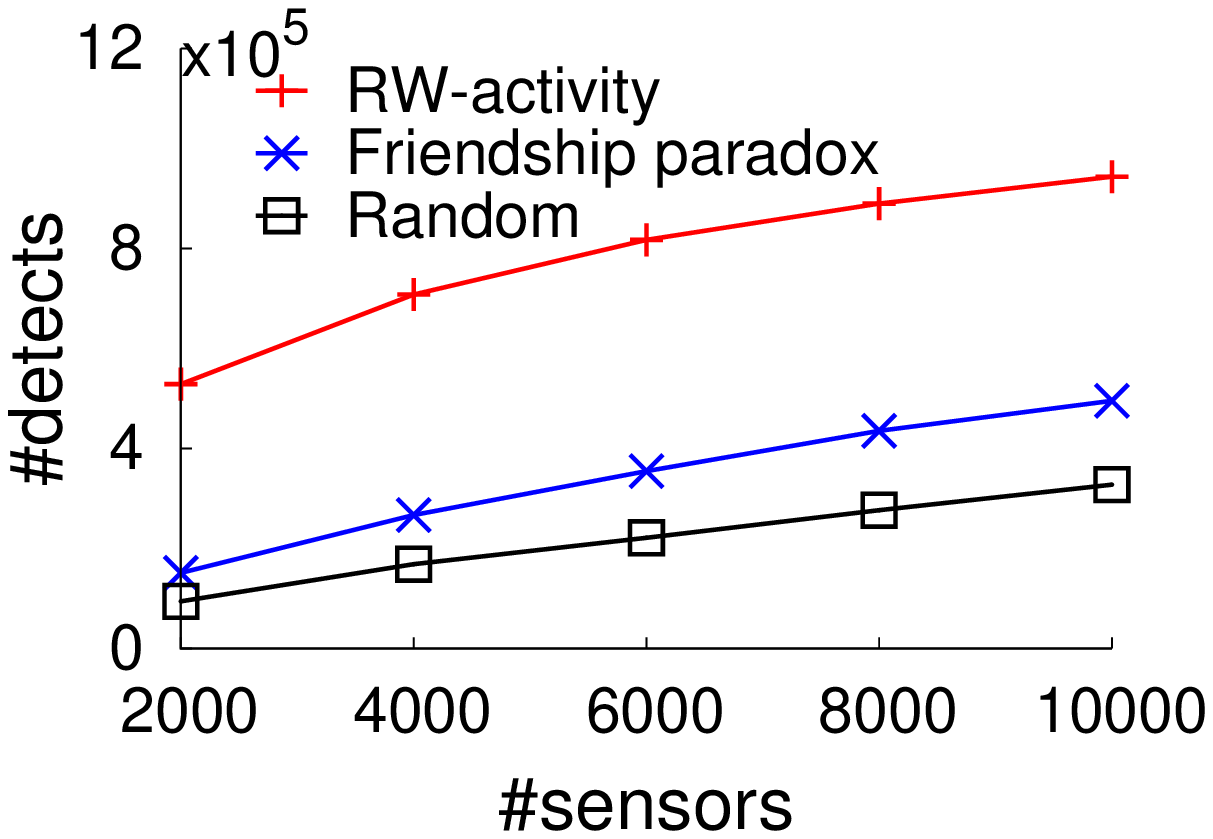}}
	\subfloat[Lead-time]{\includegraphics[width=0.5\linewidth]{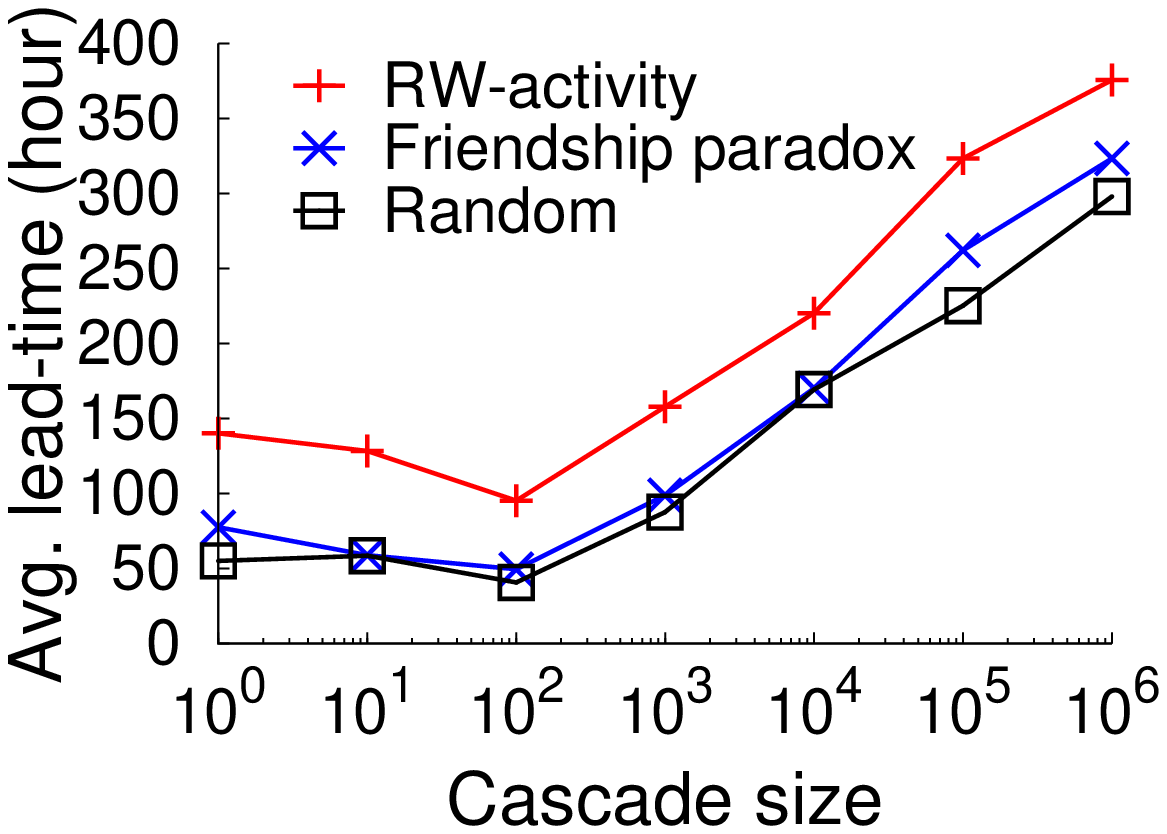}}
	\caption{Detection ability comparison.}
	\label{fig:wb_detection}
\end{figure}

We then apply the method of Alg.~\ref{alg:rw} in which random walk is biased 
by neighbors' activity and totally 50,000 Weibo accounts are collected, which form 
the candidate set. We use the posts between Jan 1, 2012 to Sep. 1, 2012 to evaluate 
the quality of a node. From these candidates we choose $B$ sensors where $B$ ranges 
from 2,000 to 10,000 using the reward function in Eq.~\eqref{eq:reward}. We also 
introduce two other measures to evaluate sensor quality: 
(a) the number of cascades sensors can detect; 
(b) the detecting lead-time, which measures the time interval 
that the sensors first detect a cascade in advance of the peak time of the cascade. 
It can be considered as the warning time for outbreaks a set of sensors can 
provide. The results are shown in Fig.~\ref{fig:wb_detection}. We observe that 
sensors obtained by Alg.~\ref{alg:rw} can discover around two to six times more 
cascades and provide earlier warning time (about two days) than the other two 
baseline methods. 

\subsubsection{\textbf{Prediction Capability Analysis}}

We study prediction in the scheme of choosing sensors based on historical data and 
testing them on future data. We want to answer: {\em Are selected sensors based on 
historical data still good for capturing future cascades?} Here we mainly compare 
the results with the method presented in \cite{Leskovec2007}, which chooses sensors 
using CELF.
% based on the historical data.
%from historical data by ``greedy on history'', i.e., choosing sensors 
%via CELF based on the 
%historical data, i.e., choosing sensors using the 
%historical data to maximize the reward function.
For our framework, we 
use Alg.~\ref{alg:rw} to select the sensor.
%first choose the candidate set by random walk, and then choose sensors from these candidates 
%(or Alg.~\ref{alg:rw}). 
%This is different from previous one. 
The Sina Weibo data 
collected by BFS used in previous section will be our ground truth data, and we 
split time into granularity of a week. 

\begin{figure}[t]
	\centering
	\subfloat[\#detects by $T_0$ ($B=100$).\label{fig:pre_frac}]
		{\includegraphics[width=0.5\linewidth]{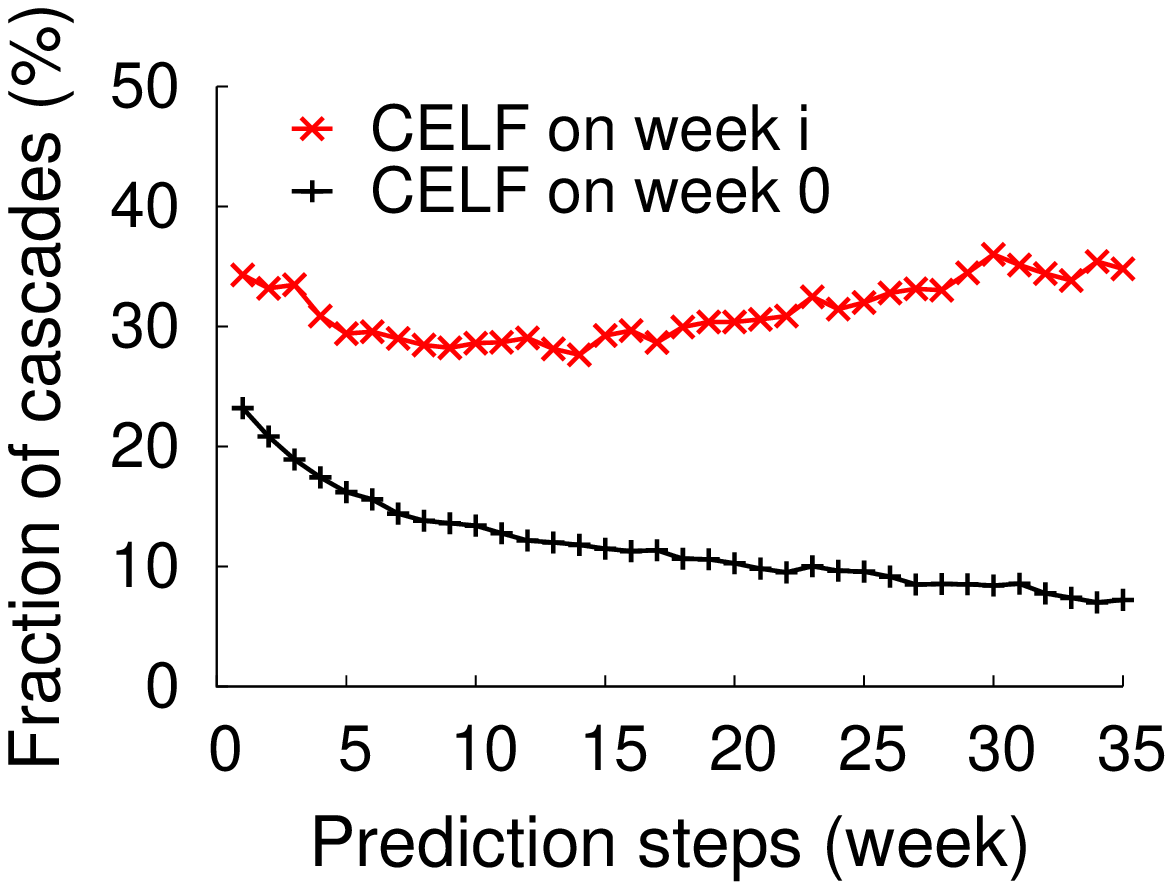}}
	\subfloat[One step prediction.\label{fig:pre_cmp}]
		{\includegraphics[width=0.5\linewidth]{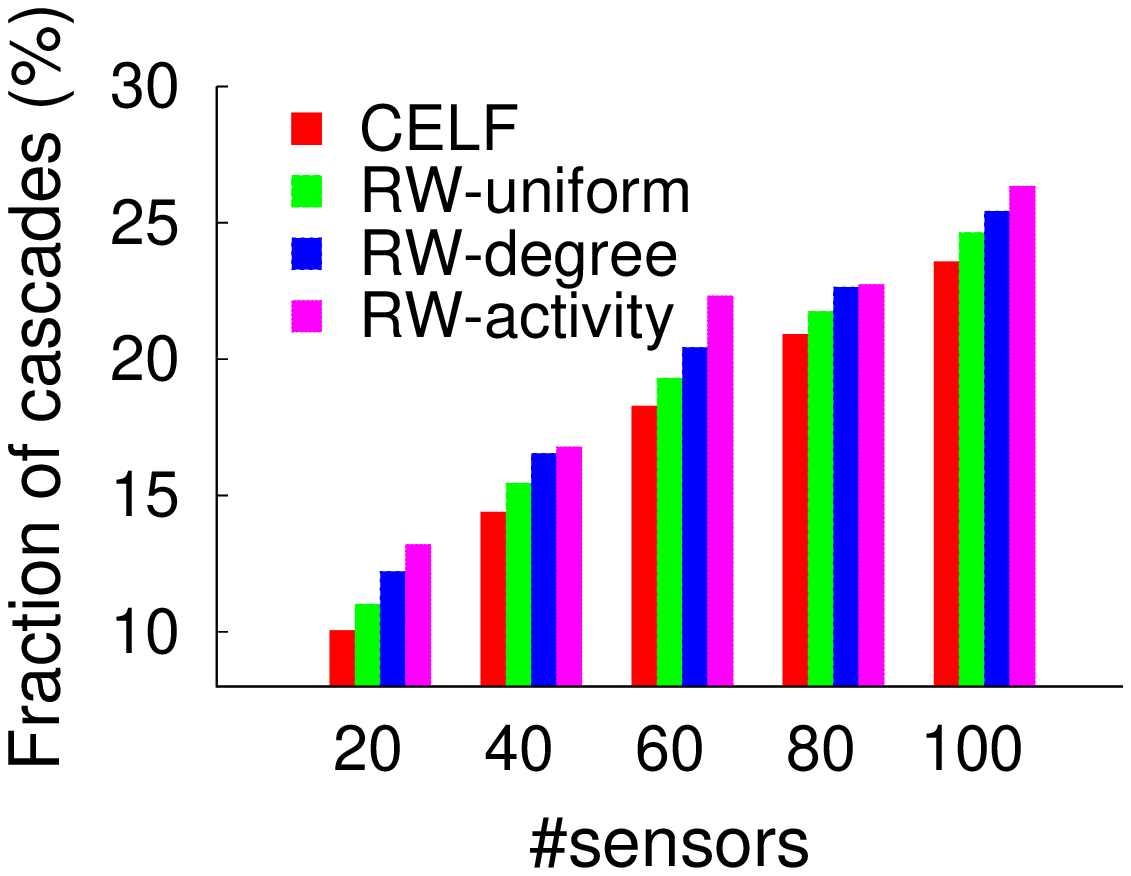}}
	\caption{Prediction ability of sensors.}
	\label{fig:wb_prediction}
\end{figure}

%First, we show the prediction results of \emph{greedy on } approach used in 
%\cite{Leskovec2007}. For the convenience of description, let $T_i,\! i\geq 0$ 
%First, we show the prediction results of \cite{Leskovec2007}. 
For the convenience of description, let $T_i, i\geq 0$ 
denote the sensors selected by the CELF algorithm using the $i^{th}$ week data.
%chosen for week $i$ using the greedy algorithm. 
Fig.~\ref{fig:pre_frac} has two curves: 
``CELF on week 0'' represents running CELF on week $0$ data only,
while ``CELF on week $i$'' represent running CELF on the $i^{th}$ week data.
%For the ``greedy on week 0'' approach, we 
%For the ``greedy on history'' approach, 
%We count the fraction of cascades 
%detected by $T_0$ for week $i, 
%i=1,2,\cdots$, which is the black curve 
%labeled by "greedy on week 0'' in Fig.~\ref{fig:pre_frac}. 
The figure depicts the fraction of cascade detected at different weeks.
The black curve corresponds to ``CELF on week 0'' while 
the red curve corresponds to ``CELF on week $i$''.
We can see that the senors in $T_0$ (or ''CELF on week 0'') is reducing its predictive capability
as time evolves.  However, if we use $T_i$ (or ``CELF on week $i$''), 
we can have a much higher predictive capability.
This implies that 
%information cascade is time varying 
social network is highly dynamic
and one needs
to execute the sensor selection algoirthm more often, rather than relying
on sensors selected based on old histrical data.
%Furthermore, we 
%also count the fraction of cascades detected by $T_i$ at week $i, i=1,2,\cdots$, 
%which is the red curve or "greedy on week $i$" in Fig.~\ref{fig:pre_frac}.

%We observe big gaps between these two curves, which means $T_0$ is loosing its 
%prediction capability as time evolves. Hence the greedy on history approach doesn't 
%generalize well on the testing data. 

\begin{comment}
The myopic of prediction is because of the existence of two types of users: type I 
users have few posts in the historical data but suddenly become active in future 
data, while type II users are active in the historical data but become inactive in 
future data. From the collected data, we find that there are more and more new 
users join in Weibo. These users are examples of type I users and they are less 
likely to be included into sensors because of the absence of historical 
information. This indicates that one has to periodically reselect sensors in 
practice because of the dynamics of user behaviors. Hence, \emph{an efficient 
algorithm is more important than an accurate algorithm in the scenario of 
prediction}. 
\end{comment}
 
Next, we study the prediction performance of our method. We choose 1,000 candidates 
by random walk biased by uniform/degree/activity respectively, from which we select 
$B$ sensors where $B$ ranges from 20 to 100, and compare the one-week prediction 
capability with aforementioned ``CELF'' approach. The result is shown in 
Fig.~\ref{fig:pre_cmp}. One interesting observation is that our approach is at least as good
as the \emph{``CELF''} approach. Random walk biased by 
users' activity is the best, which can improve about 3\% of detects on the testing data. 
Since our approach is more computationally efficient than CELF, hence, one can apply it
more often so to have a good predictive capability on information cascade, e.g.,
better than results of "CELF on week $i$".
%This indicates that our efficient approach also has some ability to overcome 
%overfit on historical data to some extent. 

\begin{figure}
	\centering
	\subfloat{\includegraphics[width=0.5\linewidth]{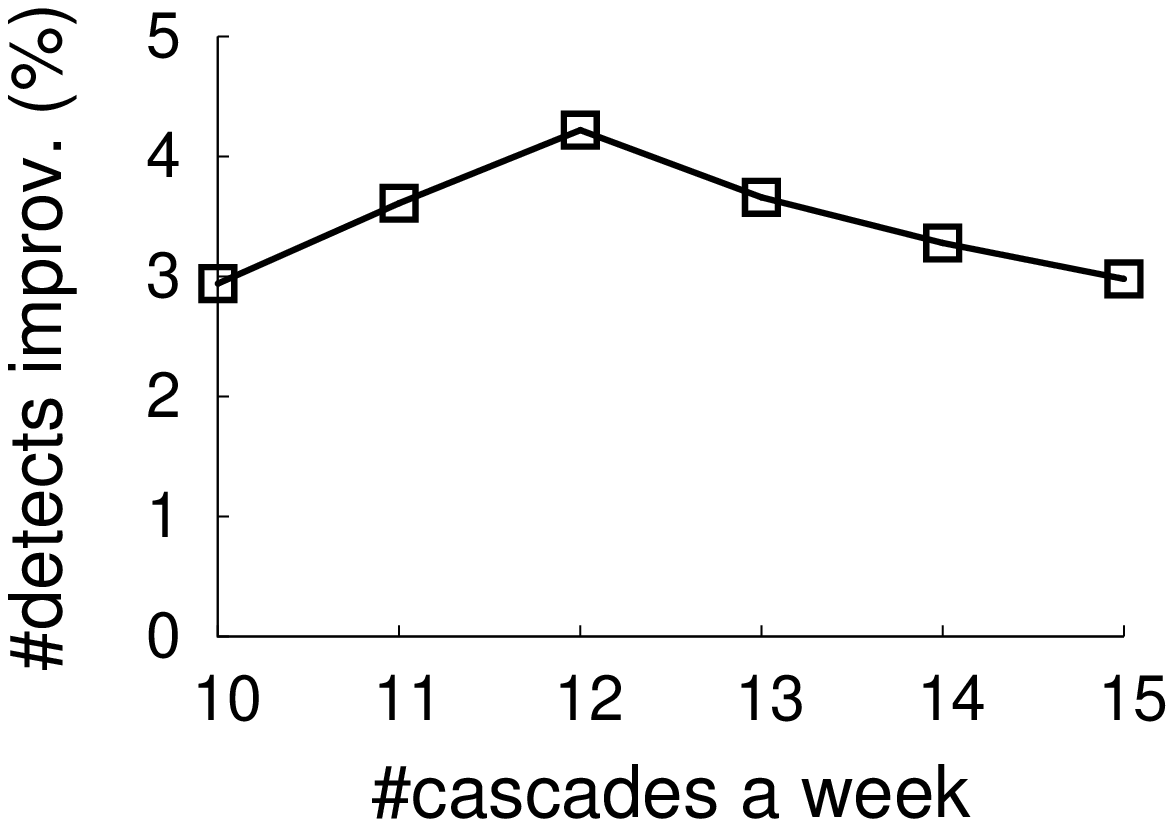}}
	\subfloat{\includegraphics[width=0.25\linewidth]{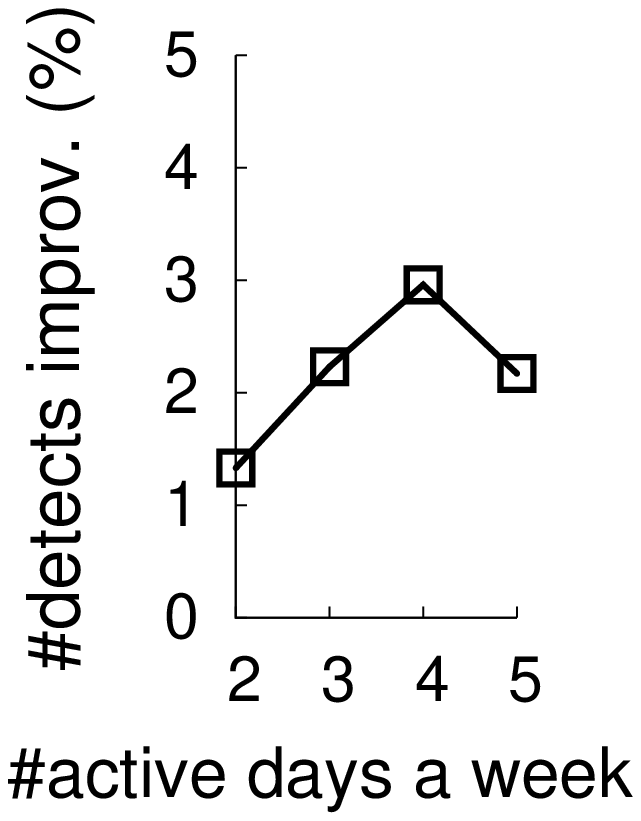}}
	\subfloat{\includegraphics[width=0.25\linewidth]{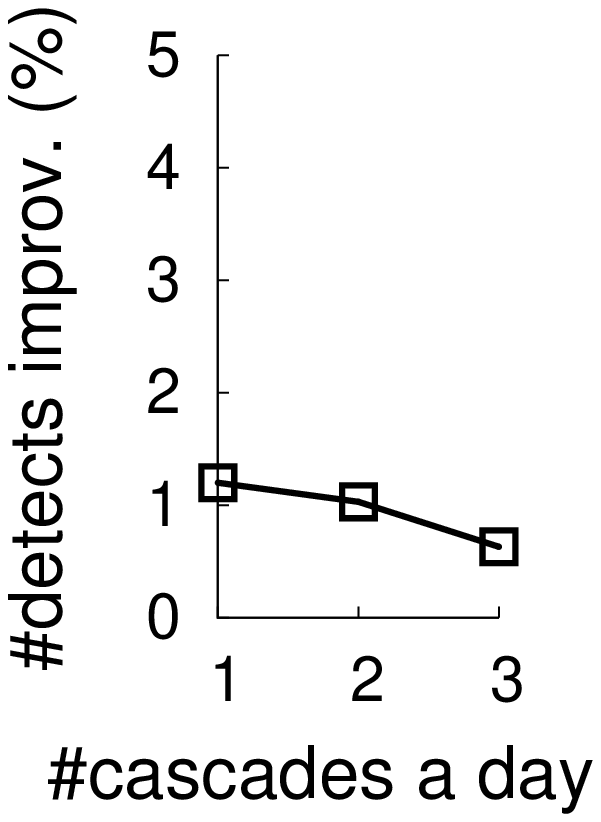}}
	\caption{One step prediction improvements relative to "greedy" 
	under different regularization rules.}\label{fig:regularize}
\end{figure}

To understand why our random walk approach has good predictive capability, 
one can consider the candidate-selecting step as a 
\emph{pruning} or \emph{regularizing} process, which are common techniques 
used to avoid overfitting of decision trees\cite[Chapter 6]{Witten2005}. 
For example, when using uniform random walk to collecting candidates, it 
prefers to select high degree nodes of the network. This can be considered as 
a rule to regularize the search scope, which can avoid choosing unimportant 
users who join into large cascades just by chance. To demonstrate this, we use 
the following regularization rules to constraint search scopes and 
then apply the CELF
%"greedy on history" approach 
to see whether there is any improvements, 
%and the results are shown in Fig.~\ref{fig:regularize}. 
 \begin{itemize}
	%\item {\em \#cascades in a week}.
        %      We set a threshold to limit our search scope 
	%      to users who have joined at least $x$ cascades in 
        %      the history data;
	\item {\em \#cascades in a week}.
              The search scope is limited 
	      to users who have joined at least $x$ cascades in 
              the history data;
	\item {\em \#active days in a week}.
              The search scope is limited to users who
	      have been active at least $x$ days in the history data;
	\item {\em \#cascades per day}. The search scope is limited to 
              users who participates at least $x$ cascades per day.
\end{itemize}
Fig.~\ref{fig:regularize} depicts the results.
We observe some improvements of prediction performance when using
different constraints, and users' activity in history is the best rule, which
is consistent with the results in Fig.~\ref{fig:pre_cmp}.
This shows that our random walk algorithm has the intrinsic property
to regularize the searching scope similar to the above regularization rules.

\section{V. Related Work} \label{sec:relatedwork}

In this section, we briefly review some related work, which are categorized 
into three groups: 

\noindent
\textbf{Influence maximization:} 
One important topic related to our work is how to identify $k$ most 
influential nodes in a social network. Domingos et al\cite{Domingos2001} first 
posed this problem in the area of marketing. Kempe et al\cite{Kempe2003} 
proved that the problem is NP-complete under both independent cascade model 
and threshold model. They also provided an approximation solution which has a 
constant factor bound. However, computing the influence spread given a seed 
set by Monte Carlo simulations is not scalable. Other 
works\cite{Chen2009,Li2012} tried to scale up the Monte Carlo simulation.
%and they are different from our sampling approach.
In our work, we do not assume any cascade or threshold model, but rather,
exploit the attributes of OSNs to determine the $B$ sensor nodes.

\noindent
\textbf{Optimal sensing:} 
Optimal sensing problems\cite{Krause2008,Leskovec2007} are closely related to
our work, which aim to find optimal observers or optimal sensor placement
strategies to measure temperature, gas concentration or detect outbreaks in
networks\cite{Leskovec2007,Christakis2010}. Our work can be considered as an
extension of Leskovec's work\cite{Leskovec2007} in which the authors proposed
the CELF approach to speedup the greedy algorithm. It is similar to the AG 
posed by Minoux\cite{Minoux1978}. There are several differences between their 
work and ours. First, we study the problem without assuming the
knowledge on the complete OSN topology. 
Second, our work aims to speedup the greedy 
algorithm via graph sampling, and that 
we can tradeoff a small loss of accuracy but 
obtain large improvement in efficiency. Furthermore,
the performance of our algorithm 
is guaranteed with high probability. The last property is important since 
AG/CELF can be as inefficiency as the greedy algorithm in the worst 
case\cite{Minoux1978}. Third, we exploit the meta information within in social 
networks. In \cite{Leskovec2007}, authors considered the problem as a discrete 
optimizing problem without considering contextual information in OSNs. In our 
experiments, we find that contextual information can be used to reduce sample 
size and improve efficiency.

\noindent
\textbf{Graph sampling and optimizing.} 
Graph sampling methods are used to measure properties of nodes and edges in 
graphs, such as degree distribution\cite{LOVASZ1993,Ribeiro2010}. 
To the best of our knowledge, this is the first work 
%However, we didn't find a work 
which uses graph sampling methods to maximize a set 
function. Lim et al\cite{Lim2011} and Maiya et al\cite{Maiya2010} both use 
different sampling methods to estimate the top $k$ largest centrality nodes in 
a graph, i.e., degree centrality, betweenness centrality, closeness centrality 
and eigenvector centrality. However these top $k$ largest centrality nodes are 
not the top $k$ optimal sensors which are defined by a set function $F(\cdot)$. 
Our work can be considered as a combination of ordinary 
optimization\cite{Ho1992} with greedy algorithm. It uses the property 
of greedy algorithm that if the reward gain at each round is bounded by factor 
$\lambda$, then the final solution obtained by the greedy algorithm is bounded 
by factor $1-1/e^\lambda$ as shown in Theorem~\ref{th:bounds}.

\section{VI. Conclusions} \label{sec:conclusion}

Many OSNs are large in scale and there is an urgent need on how to select
reliable information sources to subscribe so
one can track/detect information cascades. This can be formulated as a sensor
placement problem and previous work used heuristic greedy algorithms.
However, it is impractical to run these greedy
algorithms on a OSN composed of millions of users. Hence we propose sampling approach to 
%significantly reduce the computation complexity of 
find these good sensors.

We show that one can significantly reduce the complexity by sampling the huge
search space, and still can guarantee to have good solutions with 
high probabilistic guarantees. Hence, the sampling approach is a good method to 
trade off efficiency and accuracy. We evaluate various graph sampling approaches,
and find that random walk based sampling methods perform better than vertex
sampling based methods. This indicates that structural information of OSNs
is important and random walks are suitable for obtaining better samples, while
previous discrete optimization approaches failed to utilize this information. 
We apply our framework on Sina Weibo, and the results 
%demonstrate the advantages of sensors' detection and prediction abilities. 
demonstrate that using our algorithm, one can effectively detect
information cascade. Since our algorithm is computationally efficient,
one can execute it more often so to find new sensor nodes
to accurately predict future information cascade.

\bibliographystyle{aaai}
\bibliography{SocialSensor}

\section{Appendix}

\subsection{Proof of Theorem 1}
\begin{proof}
%\noindent
%{\bf Proof:}
By utilizing the non-decreasing property of $F$, we have
\begin{align*}
	F(OPT)-F(S_{k-1})\leq& F(OPT\cup S_{k-1})-F(S_{k-1}) \\
	=& F(OPT\backslash S_{k-1}\cup S_{k-1})-F(S_{k-1}).
\end{align*}
Assume $OPT\backslash S_{k-1}=\{z_1,\cdots,z_m\}, m\leq K$, 
and let $j=1,\cdots,m$, we have
\[
	Z_j=F(S_{k-1}\cup\{z_1,\cdots,z_j\})-F(S_{k-1}\cup\{z_1,\cdots,z_{j-1}\}).
\]
With some algebraic manipulation, we have 
\[
	F(OPT)-F(S_{k-1})\leq\sum_{j=1}^mZ_j. 
\]
Now notice that
\begin{align*}
	Z_j \leq & F(S_{k-1}\cup\{z_j\})-F(S_{k-1}) \\
	     = & \delta_{z_j}(S_{k-1}) 
	     \leq \frac{1}{\lambda}\delta_{\hat{s}_k^*}(S_{k-1}) \\
	     = &\frac{1}{\lambda}(F(S_k)-F(S_{k-1})). 
\end{align*}
Then we get an iterative formula of $F(S_k)$,
\[
	F(OPT)-F(S_{k-1})\leq\sum_{j=1}^mZ_j\leq\frac{K}{\lambda}(F(S_k)-F(S_{k-1})). 
\]
Iteratively, we can derive the following relationship,
\[
	F(S_k)\geq[1-(1-\frac{\lambda}{K})^k]F(OPT).
\]
Finally, let $k=K$. We conclude that
\[
	F(S_K)\geq[1-(1-\frac{\lambda}{K})^K]F(OPT)\geq(1-\frac{1}{e^\lambda})F(OPT).
\]
\end{proof}

\subsection{Proof of Theorem 2}
\begin{proof}
Let $y_i$ denote the number of nodes in $OPT$ that have been covered 
after the $i$-th round and $\bar{y}_i$ be its expectation. 
Further, let $q_i$ denote the probability that an uncovered 
node in $OPT$ will be covered in the $i$-th round. Then the 
expectation of $y_i$ is 
\begin{equation}
	\bar{y}_i=\bar{y}_{i-1}+q_i, 
	\label{eq:eyi}
\end{equation}
For $\alpha\%=K/|V|$ and confidence $p$, we are pretty sure that one of the $K$ 
nodes will fall in $\xi_p(\alpha)$ with probability at least $p$. Hence, the 
probability of sampling an uncovered node of $OPT$ in round $i$ is 
\[
	q_i\geq p\frac{K-\bar{y}_{i-1}}{K}. 
\]
Substituting the above equation into Eq.~\eqref{eq:eyi}, we get 
\[
	\bar{y}_i\geq\bar{y}_{i-1}+p\frac{K-\bar{y}_{i-1}}{K}.
\]
Iteratively, it can be written in the following form,
\[
\bar{y}_i\geq (1-\frac{p}{K})^i\bar{y}_0+K[1-(1-\frac{p}{K})^i]
\]
Since $\bar{y}_0=0$, we get 
\[
	\bar{y}_i\geq K[1-(1-\frac{p}{K})^i]. 
\]
After the $K$-th round, we conclude 
\[
	\E{r}=\frac{\bar{y}_K}{K}\geq 1-(1-\frac{p}{K})^K\geq 1-\frac{1}{e^p}. 
\]
\end{proof}
 
\end{document}